\documentclass[11pt]{fizik}
\usepackage{hyperref}
\hypersetup{colorlinks=true, urlcolor=blue, citecolor=blue}
\usepackage[all]{xy}
\usepackage{amsfonts,amssymb,amsmath,amsgen,amsopn,amsbsy,theorem,graphicx,epsfig}
\usepackage{amscd,bezier,latexsym,mathrsfs,enumerate,multirow} 
\usepackage[utf8]{inputenc}
\usepackage[english]{babel}
\usepackage[dvipsnames]{xcolor}
\usepackage{academicons}
\definecolor{orcidlogocol}{HTML}{A6CE39}
\usepackage[pagewise]{lineno}
\usepackage{epstopdf}
\usepackage{xspace}
\usepackage{color}
\usepackage{units}
\usepackage{slashed}
\usepackage{booktabs}
\usepackage{xcolor}
\usepackage{setspace}
\usepackage{physics}
\usepackage{nicefrac}
\usepackage{bbold}
\usepackage{accents}
\usepackage{mathtools}
\DeclareMathAlphabet{\mathcal}{OMS}{cmsy}{m}{n}

\usepackage{bm,bbm}
\usepackage{braket}
\usepackage{amsfonts}
\usepackage{framed}
\usepackage{dsfont}

\yil{}
\vol{}
\fpage{}
\lpage{}
\doi{}

\title{A Compendious Review of Majorization-Based Resource Theories: \\ Quantum Information and Quantum Thermodynamics}
\author[TORUN, PUSULUK, AND MÜSTECAPLIOĞLU]{\textbf{Gökhan TORUN$^{1}$\thanks{gtorung@gmail.com} , Onur PUSULUK$^{2}$\thanks{onur.pusuluk@gmail.com} , Özgür E. MÜSTECAPLIOĞLU$^{1,2}$\thanks{omustecap@ku.edu.tr}} \\
$^{1}$TUBITAK Research Institute for Fundamental Sciences, 41470 Gebze, Türkiye \\
$^{2}$Department of Physics, Koç University, 34450 Sarıyer, İstanbul, Türkiye \\ [1.8em]
}

\newcommand{\bc}{\begin{center}}
\newcommand{\ec}{\end{center}}

\renewcommand{\phi}{\varphi}

\begin{document}

\maketitle

\begin{abstract}
The field of quantum resource theory (QRT) has emerged as an invaluable framework for the examination of small and strongly correlated quantum systems, surpassing the boundaries imposed by traditional statistical treatments. The fundamental objective of general QRTs is to characterize these systems by precisely quantifying the level of control attainable to an experimenter. In this review article, we refrain from providing an exhaustive summary of the extensive literature on QRT. Rather, our focus centers on a specific sub-literature founded upon the theory of majorization. The primary aim is to augment our comprehension of genuine quantum phenomena manifested across diverse technological applications and incite investigations into novel resource theories encompassing multiple types of resources. Consequently, we emphasize the underlying similarities shared by various resources, including bipartite quantum entanglement, quantum coherence, and superposition, alongside informational, thermal, and generalized nonequilibrium resources.

\end{abstract}


\newpage

{\hypersetup{linkcolor={black}}
\tableofcontents}


\section{Introduction}\label{Sec:Introduction}

The framework of quantum resource theory (QRT)~\cite{Chitambar2019-QRTs, Kuroiwa2020GenQRTs} is mainly geared towards identifying, quantifying, and finding fundamental limits on harvesting the (quantum) resources that are essential for performing certain tasks in the so-called second-generation quantum technologies (QTs)~\cite{dowling2003quantum, Georgescu_2012}, including quantum computing~\cite{Hillery2016CoherenceAsResource, Chen_2021, RevModPhys.94.015004}, quantum sensing~\cite{RevModPhys.89.035002}, quantum metrology~\cite{Giovannetti2006QMetrology, Toth2014QMetrology, Friis2017QMetrology, Shlyakhov2018QMetrology}, quantum simulation~\cite{RevModPhys.86.153, RevModPhys.92.015003, PRXQuantum.2.017003}, quantum communication~\cite{Gisin2007, Chen_2021}, and quantum cryptology~\cite{Gisin2002Crypto, Portmann2022Crypto}. By focusing on a single resource required for a particular task, QRTs provide a powerful tool for developing efficient quantum protocols and devices and optimizing existing ones. The quantum advantage, which arises from the ability of quantum systems to utilize information in ways that are not achievable by classical means, has stimulated an intense effort to understand and exploit the intrinsic quantum properties. However, the full potential of QTs can only be realized if we can simultaneously identify, quantify, and manipulate different quantum resources for accomplishing a single task, which requires fusing two or more QRTs as proposed in Ref.~\cite{2023_arXiv_KK_ResourceEngines}. This is precisely the goal of general QRTs, allowing us to study the transformations between different quantum resources and fundamental limits on such quantum operations. By investigating the properties, transformations, and interplay of these resources, researchers have gained valuable insights into the fundamental limits and possibilities of QTs.

\begin{figure}[!t]
	\centering
	\includegraphics[width=1\columnwidth]{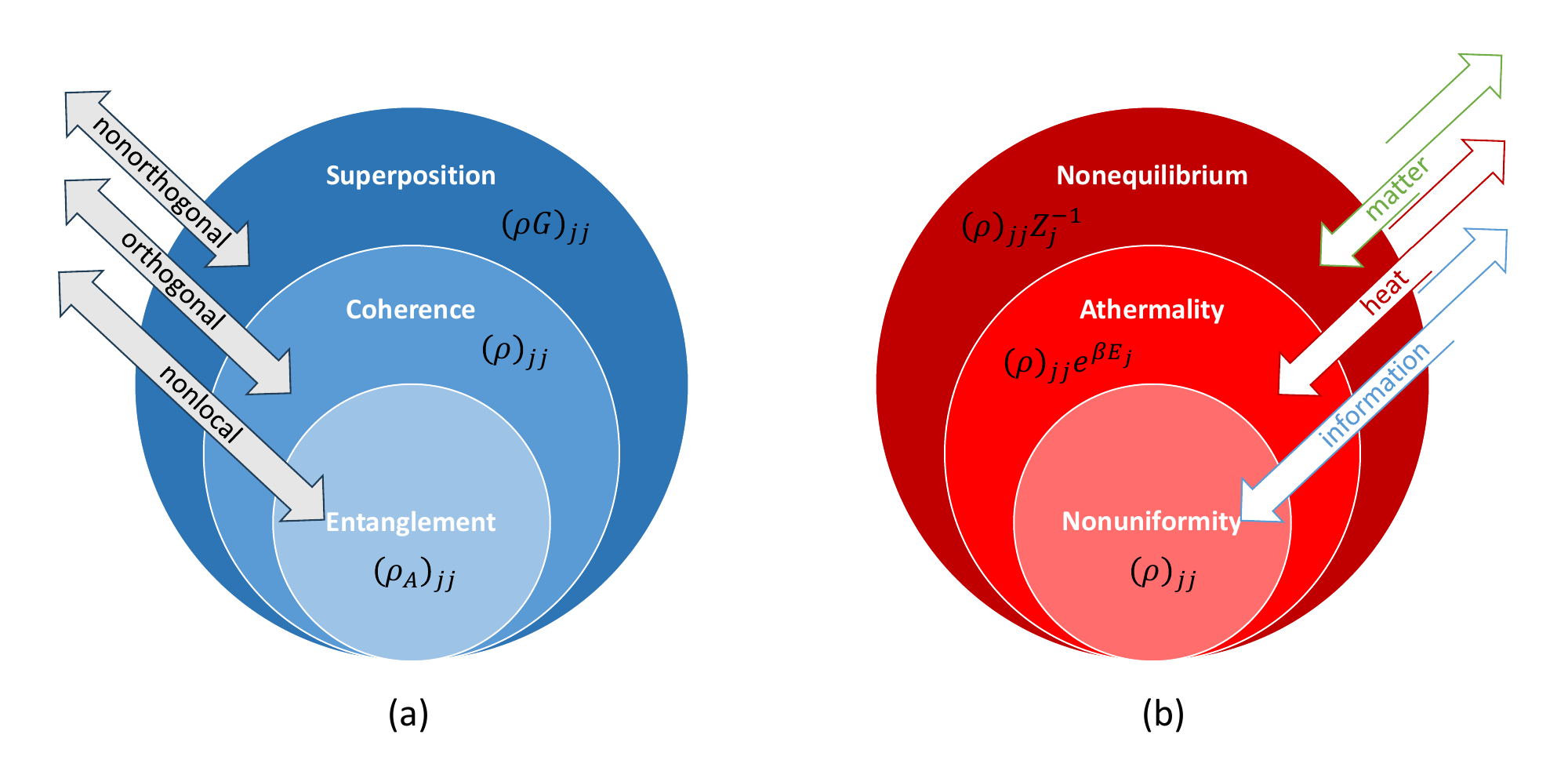}
	\caption{Visualization of the hierarchical relations among resources in (a) quantum information and (b) nonequilibrium quantum thermodynamics. The resource contained within another Venn diagram can be considered as a specific form of the other.  The arrows provide information about systems that possess these resources. The double-headed arrows represent quantum superposition, while the texts within the arrows indicate the types of states related to superposition. The merged reverse-directional arrows signify the system's interaction with its surroundings, and the texts within the arrows demonstrate the exchanged physical quantity. Under each resource, the probabilistic quantities compared in the majorization relationship, which characterizes its manipulation through relevant quantum operations, are provided. Here, the density matrix, denoted as \(\rho\), represents the quantum state, while the Gram matrix \(G\) provides information about the overlaps between the basis states in which the density matrix is expressed. \(\beta\) represents the inverse temperature, \(E\) signifies the energy, and \(Z\) denotes the partition function. The subscript \(A\) indicates that the discussion pertains to the state of subsystem A. The majorization relationship, which compares these probabilistic quantities, exhibits limitations in its application. Specifically, in (a), this relationship is limited to pure superposition states, whereas in (b), it is restricted to energy diagonal states. }
	\label{FIG:Sum}
\end{figure}

Studies over the past two decades have provided a deep understanding of the fundamental principles and mathematical tools underlying the framework of QRT. At its core, QRT distinguishes naturally a constrained set of physical operations, known as {``free''} operations, that does not produce but can consume the specific resource under consideration. Due to the limited scope of free operations, only certain physically realizable quantum states can be prepared by them, which are also referred to as free states. Any state that cannot be prepared using free operations is considered as a resource state. Consequently, QRT divides physical operations into either free or prohibited categories and likewise categorizes each quantum state as either a free state or a resource state. Free states serve as a reference point or benchmark for quantifying and comparing the resource content of other states. This versatile structure of QRT \cite{Chitambar2019-QRTs} has been effectively adapted for diverse studies, including those pertaining to quantum information in the forms of entanglement \cite{Plenio2007-QEntMeasure, Horodecki2009-QEnt} and coherence \cite{Baumgratz2014QCoherence, Streltsov2017-RTofCoh}, and disordered energy in the forms of nonuniformity \cite{2015_PhysReport_Nicole_Review, 2016_JPA_Goold_Review, 2016_ContempPhys_Anders_Review} and athermality \cite{Brandao2013RTofThermo, 2016_JPA_Goold_Review,  Ng2018, 2019_RepProgPhys_Lostaglio}. Figure \ref{FIG:Sum} visually represents the hierarchical relationships between resources in quantum information (entanglement, coherence, and superposition) and nonequilibrium quantum thermodynamics (nonuniformity, athermality, and nonequilibrium). It succinctly captures the interconnections and dependencies of these fundamental concepts in their respective domains. This paper adopts these fundamental concepts as the framework for our survey.

The highlighted aspect of QRTs here is the expansion of the set of resource states by imposing constraints on the quantum operations that an agent can perform, thereby increasing the potential applications. For instance, what makes entanglement a resource is the restriction of agents' experimental abilities to local operations and classical communication (LOCC). Otherwise, protocols like teleportation \cite{Bennett1993QTeleport} would not be possible. When we add new constraints, such as subjecting LOCC to super-selection rules, the performance of tasks where agents utilize entanglement as a resource improves, and even some previously impossible tasks become feasible~\cite{2003_PRL_SSRandLOCC, 2004_PRL_SSRandLOCC, Schuch2004SSREnt}. To fully appreciate the practical applications of QRTs, it is essential to first grasp the mathematical tools that enable its analysis. As such, the next section (Sec.~\ref{Sec:GeneralStructureofQRT}) will provide a detailed overview of these tools.

In addition to its foundational role, QRT has also found numerous applications \cite{Chitambar2019-QRTs}. The framework of QRT provides a systematic way to uncover the limits of QTs, by identifying the boundaries of what can be achieved with limited resources, providing insights into the ultimate possibilities and limitations of QTs. Quantification and characterization of these resources are helpful in understanding the nature and properties of quantum states and can guide experimentalists in assessing and benchmarking the quality of the resources they have at their disposal. For example, the characterization and quantification of quantum entanglement, which is a fundamental resource in quantum information processing (QIP), is critical for a wide range of applications, including quantum cryptography, quantum teleportation, and quantum computation. Furthermore, the concepts and techniques from QRT can be used to study the thermodynamics of quantum systems, including the resource costs and trade-offs associated with energy transformations, work extraction, and heat dissipation in quantum processes. By understanding and quantifying the resources involved, researchers can analyze and optimize the energetic aspects of quantum thermodynamic processes, contributing to the development of quantum thermodynamics as a field of study. Finally, QRT has practical applications in areas such as quantum metrology and quantum sensing, where the goal is to use quantum resources to achieve enhanced precision in measurements and sensing. QRT offers insights into resource requirements, optimization strategies, and the fundamental limits of measurement accuracy, driving progress in practical applications such as atomic clocks, gravitational wave detectors, and magnetic field sensors. Overall, the broad range of applications of QRT highlights its importance as a fundamental framework for understanding the behavior of quantum systems, and its potential for shaping the future of QTs.

Recent years have witnessed remarkable progress in the practical study of QRTs, with a particular focus on the characterization, quantification, and manipulation of different quantum resources \cite{Kuroiwa2020GenQRTs, Regula2022tightconstraints, Regula2022ProbTrQRs}. There exist comprehensive reviews in the literature that cover this topic in a manner consistent with this approach \cite{Chitambar2019-QRTs, Kuroiwa2020GenQRTs}. In this work, on the other hand, we provide an overview of the recent developments in this exciting field, with an emphasis on the practical aspects of QRT. The content of this review can be outlined as follows. Section \ref{Sec:GeneralStructureofQRT} contains the foundational principles of QRTs, encompassing the fundamental concepts of free states, resources, and free operations. Concluding this section, we outline the key properties associated with resource quantification and emphasize the role of majorization theory, which will receive further examination in subsequent discussions. Following that, we delve into the exploration of these concepts within specific resource theories, including entanglement in Sec.~\ref{Sec:RTEntanglement}, quantum coherence in Sec.~\ref{Sec:RTCoherence}, superposition in Sec.~\ref{Sec:RTSuperposition}, and resource theories of quantum thermodynamics in Sec.~\ref{Sec:ThermodynamicsApproach}. We conclude our review with an outlook in Sec.~\ref{Sec:Conclusion}.


\section{Concepts in Quantum Resource Theories}\label{Sec:GeneralStructureofQRT}

\begin{figure}[!t]
	\centering
	\includegraphics[width=.9\columnwidth]{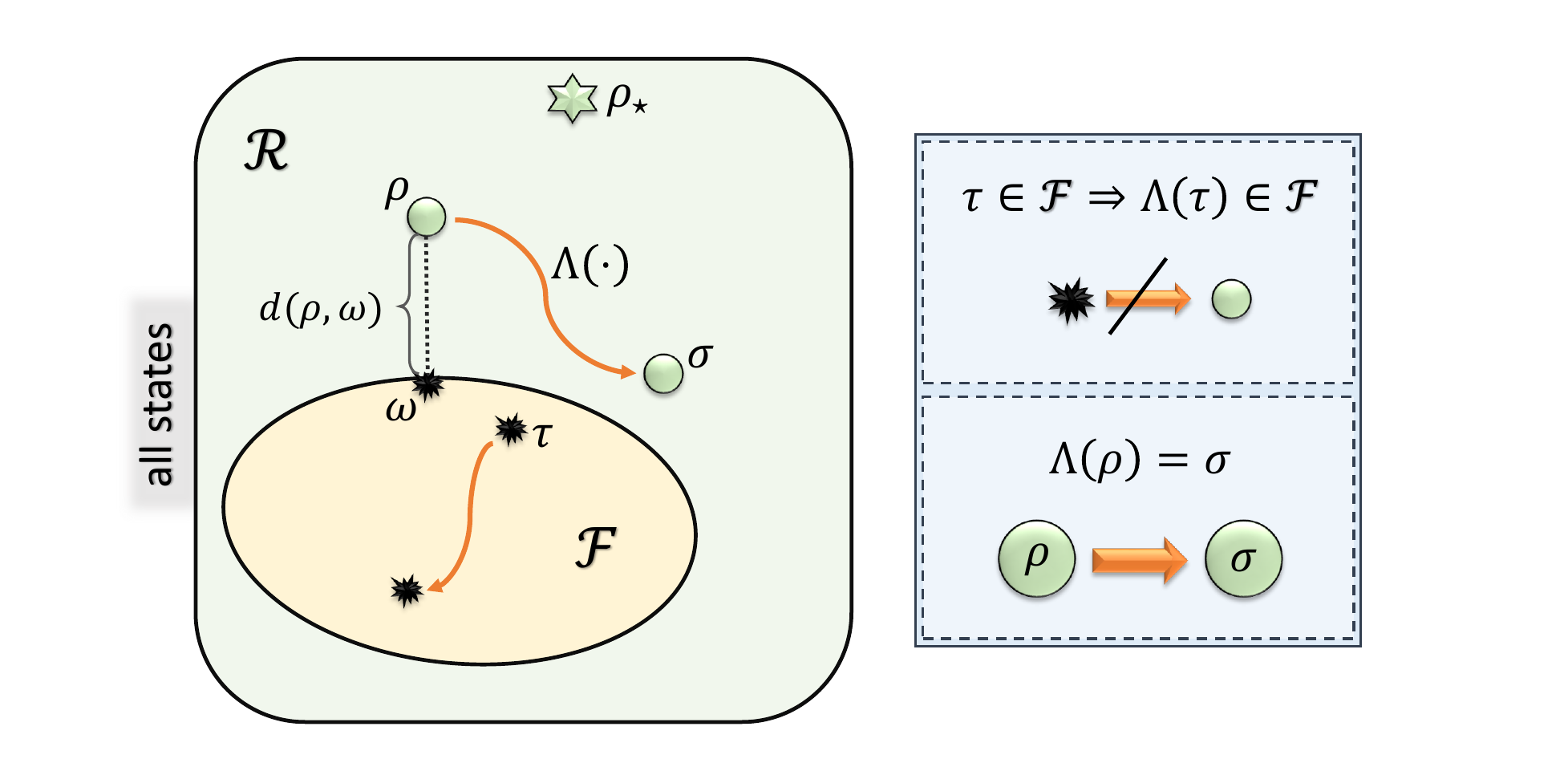}
	\caption{This figure offers a pictorial illustration that encapsulates the fundamental components central to general quantum resource theory (QRT). It visually portrays essential elements, including free states (the set they formed is denoted by \(\mathcal{F}\); e.g., \(\tau\) and \(\omega\) are two elements of this set), resource states (the set they formed is denoted by \(\mathcal{R}\); e.g., \(\rho\) and \(\sigma\) are two elements of this set), and free operations (the operation \(\Lambda \in \mathcal{O}\) that converts one resource state \(\rho\) into another resource state \(\sigma\); \(\Lambda(\rho)=\sigma\)). In addition to that, the figure depicts the golden state (\(\rho_{\star}\), representing the maximally resourceful state), the quantification of resources (\(d(\rho, \omega)\) with \(\omega\) being the free state that is closest to \(\rho\), a distance-based measure), and the rule governing resource manipulation (namely, \(\Lambda(\tau) \in \mathcal{F}\) for all \(\tau \in \mathcal{F}\), can be referred to as the \emph{golden rule} of QRTs \cite{Chitambar2019-QRTs}). Through its visually engaging representation, the figure serves as a guide, providing an enhanced understanding of the intricate dynamics and key components that define the realm of quantum resources within the expansive landscape of QRTs. The figure presented here has been sourced from Ref.~\cite{Anshu2018QuantifyingR} and has been restructured and expanded accordingly.}
	\label{FIG:ElementsofQRTs}
\end{figure}

Central to QRTs \cite{Chitambar2019-QRTs} is a structure that is shaped by three main elements: free states (the set constructed is denoted by \(\mathcal{F}\)), resource states (the set constructed is denoted by \(\mathcal{R}\)), and free operations (signified by \(\mathcal{O}\)). Of utmost importance, free states represent states that can be accessed without constraints within the given resource theory, while resource states cannot be obtained from free states through the actions of free operations; see Fig.~\ref{FIG:ElementsofQRTs}. By studying general QRTs, researchers gain insights into the nature of quantum resources and their utilization in various tasks. This analysis of resource states and the characterization of free operations deepen our understanding of the limitations, capabilities, and trade-offs involved in manipulating quantum resources, facilitating the development of practical quantum applications.

The purpose of this section is to present a highly simplified introduction to the basics of QRTs (see Sec.~III of Ref.~\cite{Chitambar2019-QRTs} for additional in-depth information). The focus of Sec.~\ref{Sec:ConstraintsFree} is on the constraints and operations within the framework of QRT and Sec.~\ref{Sec:RQuantification} discusses the characteristics of a resource quantifier. Furthermore, we dedicate the Sec.~\ref{Sec:Majorization-Based} to emphasizing the importance of majorization theory as a valuable tool within the realm of QRTs. By including these elements, we strive to facilitate readers' comprehension and foster a broad understanding of the subject matter.

\subsection{Free states, resources, and operations}\label{Sec:ConstraintsFree}

One of the central aspects of any QRT is the characterization of different classes of quantum states and operations based on their resourcefulness and the allowed transformations between them \cite{Chitambar2019-QRTs, Kuroiwa2020GenQRTs}. Our objective here is to clarify the significance of the pivotal concepts of free states, resources, and operations by exploring their respective meanings, setting the stage for the resource theory framework. Once again, we emphasize our intention to elucidate the topic in the simplest way feasible. For a more exhaustive analysis, we direct readers to refer to Ref.~\cite{Chitambar2019-QRTs}. Figure \ref{FIG:ElementsofQRTs} offers a visual elucidation of these concepts which can be summarized as follows.

\begin{itemize}
\item[(\(1\))]  \emph{Free states}:
Free states (\(\mathcal{F}\)), referred to as the \emph{unresourceful states}, are the baseline or starting point in any QRT. These states are considered abundant or readily available and do not possess any valuable resource characteristics for a specific task. For example, in the resource theory of entanglement \cite{Plenio2007-QEntMeasure, Horodecki2009-QEnt}\footnote{We note that entanglement serves as a prominent case in quantum resource theories, and therefore, its inclusion here (and below) is important for understanding the essential concepts indeed. Nonetheless, it is worth highlighting that a more extensive exploration of entanglement within the resource theory perspective will be undertaken in Sec.~\ref{Sec:RTEntanglement}}, separable (i.e., unentangled) states are regarded as free states. Overall, by characterizing free states, one establishes a reference point against which the value and usefulness of other states can be evaluated.

\item[(\(2\))]  \emph{Resources}:
Resources in quantum systems are characterized by valuable properties that enable specific tasks and exhibit distinctive features. These properties, such as entanglement \cite{Plenio2007-QEntMeasure, Horodecki2009-QEnt}, play a central role in QIP. Outside the set of free states \(\mathcal{F}\), all other states encompass resources and belong to the set of resource states (\(\mathcal{R}\)). Resource theory aims to understand, quantify, and manipulate these quantum resources to leverage their application potential. By accurately characterizing resources, we gain insights into their underlying principles and develop strategies for their manipulation and conservation. This knowledge is crucial for the effective utilization of quantum resources in diverse QIP protocols.

\item[(\(3\))]  \emph{Operations}:
General QRTs examine operations, which are physical processes that manipulate and transform quantum states. In the context of QRT, free operations are studied to understand the allowed transformations between different classes of quantum states. Basically, this analysis aims to identify which operations can convert one resource state into another and how operations can consume or generate resources. By characterizing the set of permissible operations (\(\mathcal{O}\)), we gain valuable insights into the potential and limitations of manipulating quantum resources. Put simply, the free operations \(\mathcal{O}\) are incapable of transforming any state within \(\mathcal{F}\) into a state that lies outside the set \(\mathcal{F}\). The study of free operations in general QRTs focuses on determining the transformative capabilities of these operations, their impact on resources, and the required operational resources for their implementation \cite{Chitambar2019-QRTs, Kuroiwa2020GenQRTs}.
\end{itemize}

A resource theory can be applied to analyze various scenarios by classifying actions as either free or prohibited, and subsequently exploring the possibilities afforded by the free operations. Within this framework, certain objects become regarded as resources, as they cannot be generated in the given setting. Envision the following illustrative example, which can contribute to the understanding of these basic concepts of QRTs introduced above (the introductory section of Ref.~\cite{Chitambar2019-QRTs} also commences by illustrating an example within the given context). Imagine a student preparing for an exam with limited resources at their disposal. They are only permitted to use a textbook and their own notes, while access to the internet or additional reference materials is prohibited. In this context, the student views internet access and supplementary resources as valuable resources since they are unavailable and cannot be utilized for exam preparation. By providing a concrete example, the concept of ``free'' can be more effectively elucidated, facilitating a better understanding of its meaning and implications.

To summarize, free states refer to quantum states that can be prepared without requiring any external resources, typically through local operations. On the other hand, resource states exhibit nonlocal properties that render them valuable for specific quantum tasks. Free operations, which are operations that can be performed without expending external resources, maintain the set of free states throughout their execution. In Sec.~\ref{Sec:Information-theoretic}, we will explore the phenomena of entanglement (Sec.~\ref{Sec:RTEntanglement}), quantum coherence (Sec.~\ref{Sec:RTCoherence}), and superposition (Sec.~\ref{Sec:RTSuperposition}) respectively, providing an understanding of the interplay between the triple \(\{\mathcal{F}, \mathcal{R}, \mathcal{O}\}\). As the next phase, Sec.~\ref{Sec:ThermodynamicsApproach} provides an exposition on the subject matter, specifically focusing on nonequilibrium quantum thermodynamics.

The relationship between free states \(\mathcal{F}\), resource states \(\mathcal{R}\), and free operations \(\mathcal{O}\) serves as the foundation for quantifying quantum resources. In other words, by establishing resource monotones, measures that quantify the amount of a particular resource present in a state, we gain a means to classify, compare, and assess the hierarchy of resource states \cite{Plenio2007-QEntMeasure}. Thus, the subsequent section (Sec.~\ref{Sec:RQuantification}) will explore the methodologies employed for precisely measuring and characterizing quantum resources within the framework of general QRTs.


\subsection{Quantifying quantum resources}\label{Sec:RQuantification}

Quantum resource quantifiers assume a significant role within the framework of QRTs by enabling a systematic and quantitative evaluation of the resourcefulness exhibited by quantum systems and operations. These quantifiers serve as essential tools for characterizing, comparing, and assessing the availability and distribution of resources across diverse quantum scenarios. More specifically,  the knowledge derived from resource quantification contributes to the development of more efficient quantum algorithms, protocols, and technologies, thereby pushing the boundaries of what can be achieved in the quantum realm \cite{Chitambar2019-QRTs}.

To establish a resource measure with rigor, additional structural elements are necessary, typically expressed through a set of axioms. Systematic research endeavors have been dedicated to the extensive characterization and computational analysis of resource measures in the context of general QRTs \cite{ZWen2017RDesMaps, Bromley2018AccBoundsQR, Regula2018ConvexGeomtryQR, Designolle2019Incompatibility, Uola2019QuanifyingQR, Ducuara2020WeightBased, Tendick2023distancebased}. For instance, Regula \cite{Regula2018ConvexGeomtryQR} introduced a unified framework for mathematically characterizing various measures of general quantum resources. This framework enables a systematic definition of faithful quantifiers within any convex quantum resource theory. In Sec.~VI of Ref.~\cite{Chitambar2019-QRTs}, Chitambar and Gour initiated their investigation through an axiomatic approach, whereby they sought to identify a set of essential and desirable properties that must be upheld by any resource measure employed in the realm of QRTs. Subsequently, they undertook a comprehensive examination of various categories of specific resource measures that can be effectively employed in the broader analysis of QRTs. In accordance with Ref.~\cite{Chitambar2019-QRTs}, the defining characteristics of a proper resource quantifier can be elucidated as follows:

\begin{itemize}
  \item[(A1)] \textbf{Non-negativity}: A valid resource measure \(\mathcal{M}\) should assign a value of zero to the set of free states \(\mathcal{F}\), which represent states that do not possess any resourcefulness. In general
\begin{eqnarray}\label{Nonnegativity}
\mathcal{M}(\rho) \geq 0,
\end{eqnarray}
with equality holding if and only if \(\rho\in \mathcal{F}\).

  \item[(A2)] \textbf{Monotonicity}: The resource quantifier \(\mathcal{M}\) should exhibit monotonic behavior, meaning that it does not increase under the action of free operations \(\Lambda(\cdot)\). This axiom ensures that the measure reflects the intuitively desirable notion that resourcefulness should not be increased by using the corresponding free operations. That is,
\begin{eqnarray}\label{Monotonicity}
\mathcal{M}\big(\Lambda(\rho)\big)  \leq \mathcal{M}(\rho).
\end{eqnarray}

 \item[(A3)] \textbf{Convexity}: The measure \(\mathcal{M}\) should be a convex function, meaning that it assigns a higher value to convex combinations of resource states compared to non-convex combinations. Convexity captures the idea that combining multiple resources should lead to a more valuable composite resource. That is,
\begin{eqnarray}\label{Convexity}
\sum_{i} p_i \mathcal{M}(\rho_i) \geq  \mathcal{M}\left(\sum_{i} p_i \rho_i\right).
\end{eqnarray}

  \item[(A4)] \textbf{Subadditivity}: This axiom reflects the behavior of the measure when applied to composite systems. It implies that the resource measure of a composite system should be less than or equal to the sum of the measures of its individual subsystems. Namely, for all \(\rho\) and \(\sigma\), \(\mathcal{M}\) is additive under tensor products:
\begin{eqnarray}\label{Subadditivity}
\mathcal{M}({\rho}\otimes{\sigma})  \leq \mathcal{M}({\rho}) + \mathcal{M}({\sigma}).
\end{eqnarray}

  \item[(A5)] \textbf{Asymptotic Continuity}: This axiom pertains to the behavior of the measure in the asymptotic limit. It requires that the measure remains continuous as the size of the system increases, allowing for a smooth transition from finite to infinite-dimensional systems.
\end{itemize}

Resource quantification plays a crucial role in analyzing the efficiency and limitations of quantum protocols. For a given initial state \(\rho\) and a target state \(\sigma\), by comparing resource measures before and after transformations, we can gain valuable insights into the effectiveness of operations in manipulating and enhancing quantum resources. This approach optimizes resource manipulation strategies, leading to improved resource utilization. In our discussion of different resource theories, we will explore various resource quantifiers and their associated properties. Moreover, readers are encouraged to refer to \cite{Chitambar2019-QRTs} for a more extensive treatment of the subject matter.


\subsection{Majorization as a tool for resource theory}\label{Sec:Majorization-Based}

Majorization captures the intuitive idea that the components of vector \(\mathbf{x}\) are typically ``less spread out'' or ``more nearly equal'' than those of vector \(\mathbf{y}\). This can be expressed by stating that \(\mathbf{x}\) is majorized by \(\mathbf{y}\), written \(\mathbf{x} \prec \mathbf{y}\). Marshall \emph{et al}.~\cite{Marshall1979InequalitiesMAJ} provide historical insights into majorization and extensively cover its theoretical foundations and applications. Importantly, with broad utility across mathematics \cite{Marshall1979InequalitiesMAJ}, economics \cite{Saposnik1993MAJECO, Arnold2018MAJECO, Kleiner2021MAJECO}, statistics, and quantum physics \cite{Buscemi2012QSDecision, buscemi2015fully, 2018_NatureComm_qMajor}, majorization theory serves as a powerful tool for comparing component distributions, analyzing inequalities \cite{Marshall1979InequalitiesMAJ, Bhatia1996Matrix, Cicalese2002}, optimizing systems, and exploring entanglement in quantum mechanics \cite{Nielsen1999MAJ, NielsenVidal2001Majorization, Nielsen2002MAJIntroduction}. Its elegant framework facilitates the understanding of various phenomena and drives advancements in multiple disciplines.

The majorization theory is a mathematical concept \cite{Bhatia1996Matrix} used for comparing two real vectors $\mathbf{p}$ and \(\mathbf{q}\), where \(\mathbf{p} = (p_1, p_2, \dots, p_n)\) and \(\mathbf{q} = (q_1, q_2, \dots, q_n)\) are the respective elements of the vectors. Specifically, given two vectors \(\mathbf{p}\) and \(\mathbf{q}\), with \(\mathbf{p} \prec \mathbf{q}\) denoting the majorization relationship between them (i.e., \(\mathbf{p}\) is majorized by \(\mathbf{q}\)), the ensuing condition applies: For every value of \(k\) ranging from 1 to \(n-1\), the sum of the \(k\) largest elements in \(\{p_i\}\) is less than or equal to the sum of the \(k\) largest elements in \(\{q_i\}\). Mathematically, it can be expressed as
\begin{eqnarray}\label{Def:Majorization}
\sum_{i=1}^k p_i^{\downarrow} \leq \sum_{i=1}^k q_i^{\downarrow} \quad \text{for all} \;  k=1, 2, \dots, n-1,
\end{eqnarray}
with equality holding when \(k\) is equal to \(n\). Here, \(p_i^{\downarrow}\) and \(q_i^{\downarrow}\) denote the \(i\)-th largest elements in the vectors \(\mathbf{p}\) and \(\mathbf{q}\), respectively, when sorted in non-increasing order. To recap, \(\mathbf{p}\) is majorized by \(\mathbf{q}\), if the cumulative sums of elements in \(\mathbf{p}\) are always less than or equal to the corresponding sums in \(\mathbf{q}\) when considering the largest \(k\) elements.  For further insights into majorization theory, Ref.~\cite{Marshall1979InequalitiesMAJ} offers a wealth of information.

Majorization plays a fundamental role in quantifying and comparing entanglement levels in bipartite quantum states \cite{Nielsen1999MAJ, NielsenVidal2001Majorization, Nielsen2002MAJIntroduction}. Clearly, Nielsen's study \cite{Nielsen1999MAJ} stands as a seminal contribution that introduced and catalyzed subsequent investigations into the application and implications of this theory. This pioneering work \cite{Nielsen1999MAJ} provided the bedrock for a multitude of subsequent studies, establishing a robust framework for further exploration and advancement in the field. As demonstrated in Ref.~\cite{Nielsen1999MAJ}, for a bipartite quantum state \(\ket{\psi}_{AB}\), the possibility of transforming it into another bipartite quantum state \(\ket{\phi}_{AB}\) can be determined by considering majorization criteria. Specifically, let \(\rho_A\) (\(\sigma_A\)) be the state of the first subsystem, which is obtained by performing a partial trace over the second subsystem, that is, \(\rho_A = \text{tr}_B(\rho)\) (\(\sigma_A = \text{tr}_B(\sigma)\)), where \(\rho=\ket{\psi}_{AB}\bra{\psi}\) (\(\sigma=\ket{\phi}_{AB}\bra{\phi}\)). The vector of eigenvalues corresponding to \(\rho_A\) (\(\sigma_A\)), arranged in non-increasing order, is denoted by \(\lambda(\psi)\) (\(\lambda(\phi)\)). Then, \(\ket{\psi}_{AB}\) can be transformed into \(\ket{\phi}_{AB}\) through free operations if and only if \(\lambda(\psi)\) is majorized by \(\lambda(\phi)\). In shorthand, this condition is expressed as \(\ket{\psi}_{AB} \rightarrow \ket{\phi}_{AB}\) if and only if  \(\lambda(\psi) \prec \lambda(\phi)\) \cite{Nielsen1999MAJ}. Overall, majorization forms the basis for quantitatively analyzing entanglement in bipartite pure quantum systems \cite{Nielsen1999MAJ, NielsenVidal2001Majorization, Nielsen2002MAJIntroduction}, and specific example(s) will be discussed in Section~\ref{Sec:RTEntanglement} to probe this topic further.

In addition to these, the concept of majorization serves as a valuable tool in numerous other studies and investigations. For instance, currently, how majorization can be used to analyze the performance of a quantum Otto engine in the quasistatic regime was investigated in Ref.~\cite{Sachin2023Spin-basdMaj}. Buscemi and Gour \cite{Buscemi2017LorenzCurve} explored the role of majorization and its variants in physical resource theories, introduced a unifying framework using quantum relative Lorenz curves, and assessed transformations between quantum states, particularly in the context of resource theory of athermality. In a distinct study presented by Buscemi \emph{et al.} \cite{Buscemi2019information} a simple sufficient condition, based on one-shot relative entropies and quantum relative majorization, was established to decide the existence of a quantum channel transforming one pair of quantum states into another, with implications on the rate of transformation and applications to resource theories of athermality and coherence. As a result of its broad applicability and robust theoretical underpinnings, majorization has emerged as an indispensable and fundamental tool in a multitude of research investigations.



\section{Quantum Information}\label{Sec:Information-theoretic}


Embracing a wide spectrum of concepts \cite{Vedral1997QuantifyingEnt, Gour2008RTReferenceFrame, Horodecki2013QRTs, Grudka2014QContext, Veitch2014RTStab, Brandao2015RevQRTs, Brando2015TheSecondLaw, de_Vicente_2016, COECKE201659, Oppenheim2017WorkandHeat, GGour2017SingleShot, Kaifeng2017CohMeasure, Streltsov2017RTCDistributed, ZWen2017RDesMaps, Howard2017AppRTofMagic, Lami2018GaussianRT, Ng2018, Yadin2018OperQRTsNonclassicality, Zhuang2018RTNonGaussian, Takagi2019SubDics, Oszmaniec2019operational, Takagi2019GenQRTsBeyond, Wang_2019Magic, Liu2019OneShotQRT, Wang2019RTAsymDistiQChan, Wang2019RTAsymmetryic, Takagi2020AppRTofChan, LLu2020QuantQC, Buscemi2020CompleteRT, Liu2020OpRTs, Wolfe2020quantifyingbell, Zhou2020StateTrRTs, Kristjansson_2020, Khanian2022, Halpern2022RTComplexity, Ferrari2022, Zjawin2023quantifyingepr, Garcia2023, Buscemi2023unifyingdifferent, wu2023resource}, the explored approaches in QRTs encompass pivotal exemplifications like entanglement \cite{Plenio2007-QEntMeasure, Horodecki2009-QEnt}, quantum coherence \cite{Baumgratz2014QCoherence, de_Vicente_2016, Kaifeng2017CohMeasure}, and superposition \cite{Theurer2017RTS}. Over the past two decades, the field of QRT has experienced rapid growth, surpassing the point where it can be comprehensively captured in a single review. Therefore, we recognize the presence of previous reviews that concentrated on specific resource theories \cite{Plenio2007-QEntMeasure, Horodecki2009-QEnt, Streltsov2017-RTofCoh, Chitambar2019-QRTs}, offering valuable guidance through our survey. Our intention here is to provide a brief analysis of entanglement (Sec.~\ref{Sec:RTEntanglement}), followed by a critical exploration of coherence (Sec.~\ref{Sec:RTCoherence}) and superposition (Sec.~\ref{Sec:RTSuperposition}) theories. By establishing their conceptual affinities, this study seeks to foster a deeper understanding of the interrelationships between these theoretical frameworks (see Fig.~\ref{FIG:Sum}, part (a)). As we proceed to the next section, we find the presented details on entanglement and quantum coherence to be sufficiently elucidating. It is worth reiterating that our primary focus pertains to the theory of superposition.

\subsection{Resource Theory of Entanglement}\label{Sec:RTEntanglement}

The theory of entanglement stands out as an illustrious and highly regarded paradigm within the realm of QRTs. Over the last two decades, extending to the current date, extensive study and analysis have been dedicated to this subject \cite{Vedral1997QuantifyingEnt, Cleve1997QEnt, Karol2001DynamicsEntang, Wei2003GeometricMeasureEnt, Plenio2007-QEntMeasure, Horodecki2009-QEnt, Vedral2014QEnt, Erhard2020QEnt, Duarte2022, Lami2023NoSecondLaw}. In their comprehensive review of entanglement measures, Plenio and Virmani \cite{Plenio2007-QEntMeasure} concentrated on specific topics (i.e., tools for quantifying entanglement), providing a thorough exploration of each. Shortly thereafter, Horodecki \emph{et al.} \cite{Horodecki2009-QEnt} undertook a detailed review, delving into various facets of quantum entanglement, thus offering an authoritative analysis of the subject. Undoubtedly, as the most celebrated exemplar of general QRT, the study of entanglement unravels profound insights into the intricate nature of quantum correlations \cite{Adesso2016QCorre, Fanchini2017LecturesQCorre} and offers invaluable tools for manipulating, characterizing, and harnessing this captivating resource in a myriad of information-theoretic tasks. Given the availability of extensive reviews on the resource theory perspective of entanglement \cite{Plenio2007-QEntMeasure, Horodecki2009-QEnt}, the objective of this section is to present a concise outline of quantum entanglement, serving as a reference point for exploring a wide range of approaches in QRT. It is worth noting that our explanations will follow a similar flow to the one presented in Ref.~\cite{Chitambar2019-QRTs}.

The theory of entanglement encompasses a physical scenario where spatially separated parties (e.g., let us consider two parties Alice and Bob, see Fig.~1 in Ref.~\cite{Plenio2007-QEntMeasure}) have the ability to freely exchange classical information while all quantum information is processed locally using completely positive and trace-preserving (CPTP) maps\footnote{A map \(\mathcal{E}\) is said to be \textit{trace preserving} (TP) if  \(\textbf{Tr}(\mathcal{E}(\rho))=\text{Tr}(\rho)\). One also requires that \(\mathcal{E}\) maps positive operators to positive operators. Also, a map \(\mathcal{E}\) is said to be \textit{completely positive} (CP) if \(\mathcal{E} \otimes I_d\) is positive for all \(d \in \mathds{N}\), where \(I_d\) denotes the \(d \times d\) identity operator. Thus, a map satisfying these two properties is said to be \textit{completely positive trace preserving} (CPTP).} on individual subsystems. The class of local operations and classical communication (LOCC) corresponds to the set of operations that can be implemented within this restriction, representing the free operations in the context of quantum entanglement, as they do not generate or consume entanglement. Importantly, the sequential nature of LOCC ensures that they preserve the separability of states, i.e., applying LOCC to separable states will always result in separable states \cite{Horodecki2009-QEnt}.


Separable states --- free states in the context of quantum entanglement --- are quantum states expressed as a convex combination of product states, characterized by the absence of entanglement. For the composite system with two parties, a separable state (i.e., unentangled) can be written as
\begin{equation}\label{Def:SepStates}
\rho_{\text{sep}} = \sum_i p_i \rho_i^{A} \otimes \rho_i^{B}.
\end{equation}
Here, \(\{p_i \geq 0\}_{i}\) represents the probabilities associated with each product state, and \(\rho_i^{A}\) and \(\rho_i^{B}\) are the density matrices of the individual subsystems of Alice and Bob, respectively. The concept of separable states extends naturally to \(N\)-party systems, allowing for a description of a separable state as a convex combination of product states for each subsystem. That is, in the case of an \(N\)-party composite system, a separable state can be expressed as \(\sum_i q_i \rho_{i}^{(1)} \otimes \rho_{i}^{(2)} \otimes \ldots \otimes \rho_{i}^{(N)}\), where \(\{q_i \geq 0\}_{i}\) denotes the probabilities associated with each product state, and \(\rho_{i}^{(k)}\) for \(k=1, 2, \dots, N\) represents the density matrix of the \(k\)-th subsystem (i.e., capturing its local properties). This encapsulates the separability of the state by expressing it as a combination of local density matrices for each party, reflecting the absence of entanglement between the subsystems.

The definition of separable states, as described by Eq.~\eqref{Def:SepStates} for bipartite systems, appears obvious upon careful consideration. However, it is essential to acknowledge that more intricate scenarios can arise, leading to a broader range of possibilities \cite{Horodecki2009-QEnt}. In other words, the study of separable states has led to the exploration of various families with distinct properties. For instance, one such family is fully separable states, where each subsystem exists in a pure state, allowing for a product state representation without entanglement. Another notable family is biseparable states, which can be partitioned into two subsystems such that each subsystem is separable. This characterization enables a deeper understanding of entanglement by focusing on the separability of subsystems within larger quantum systems. Additionally, certain classes of mixed states, such as Werner states \cite{Werner1989, Azuma2006Werner} and isotropic states \cite{Horodecki1999Isotropic, Terhal2000Isotropic}, demonstrate separability under specific conditions. For instance, the two-qubit Werner state \cite{Werner1989} is described by the density operator \(\rho_{W_z} = z\ket{\Psi^{-}}\bra{\Psi^{-}} + ([1-z]/{4}) \mathds{1}_2 \otimes \mathds{1}_2\), where \(\mathds{1}_2\) denotes the two-dimensional identity operator and \(\ket{\Psi^{-}} = ({\ket{01} - \ket{10}})/{\sqrt{2}}\) represents the singlet state. Obviously, the Werner state \(\rho_{W_z}\) is characterized by the parameter \(z \in [0,1]\). Specifically, for \(z \in \left(\frac{1}{3},1\right]\), the Werner state \(\rho_{W_z}\) is entangled; however, for \(z \in \left[0, \frac{1}{3}\right]\), the Werner state \(\rho_{W_z}\) is separable \cite{Werner1989}.

On the other hand, resource states are quantum states that possess nontrivial entanglement and are valuable for performing QIP tasks. These states exhibit correlations between subsystems that cannot be described using classical resources alone, enabling tasks such as quantum teleportation \cite{Bennett1993QTeleport}, superdense coding \cite{Bennett1992DenseCoding}, and quantum key-distribution \cite{Ekert1991KeyDist}. In the case of quantum entanglement, several families of resource states have been extensively studied due to their significance \cite{Horodecki2009-QEnt}. These include maximally entangled states, such as Bell states \(\ket{\Psi^{\mp}} = ({\ket{01} \mp \ket{10}})/{\sqrt{2}}\) and \(\ket{\Phi^{\mp}} = ({\ket{01} \mp \ket{10}})/{\sqrt{2}}\) \cite{Plenio2007-QEntMeasure, Horodecki2009-QEnt}, and Greenberger-Horne-Zeilinger states \(\ket{\mathrm{GHZ}}=(\ket{000}+\ket{111})/{\sqrt{2}}\) \cite{Dur2000ThreeQubitGHZ}, which exhibit the highest degree of entanglement and enable various applications in quantum communication and computation \cite{Ghosh2002TeleportGHZ}.

We now embark upon the examination of entanglement measures which provide a quantitative assessment of the amount of entanglement present in a given state. Entanglement measures possess important properties discussed in Sec.\ref{Sec:RQuantification}. They are non-negative and vanish for separable states, reflecting the absence of entanglement in such states. Mathematically, an entanglement measure \(\mathcal{E}\) satisfies \(\mathcal{E}(\rho) \geq 0\) for all quantum states \(\rho\), and \(\mathcal{E}(\rho) = 0\) if and only if \(\rho\) is separable. Entanglement measures are also monotonic under LOCC, meaning that the (average) amount of entanglement cannot increase under free operations. Additionally, entanglement measures are convex, reflecting the fact that mixing entangled states cannot increase the overall entanglement. Mathematically, for any convex combination \(p\rho + (1-p)\sigma\) of quantum states \(\rho\) and \(\sigma\), the entanglement measure satisfies \(\mathcal{E}\big(p\rho + (1-p)\sigma\big) \leq p\mathcal{E}(\rho) + (1-p)\mathcal{E}(\sigma)\). Various types of entanglement measures have been developed to capture different aspects of quantum entanglement \cite{Plenio2007-QEntMeasure} and can be used to compare and classify different resource states based on their entanglement content.

Entanglement measures play a crucial role in the theory of entanglement \cite{Plenio2007-QEntMeasure, Horodecki2009-QEnt}, as they allow for the comparison, classification, and manipulation of different entangled states. A variety of measures have been developed to quantify entanglement \cite{Plenio2007-QEntMeasure}, and among them, three notable ones can be succinctly summarized as follows. First, the entanglement entropy \cite{Plenio2007-QEntMeasure}
\begin{eqnarray}\label{EntEntropy}
E(\rho) = -\text{Tr}(\rho\log\rho),
\end{eqnarray}
which is based on the von Neumann entropy of the reduced density matrix. Second, the concurrence \(C(\rho)\), which quantifies entanglement in bipartite systems and is defined as
\begin{eqnarray}\label{Concurrence}
C(\rho) = \max\{0, \lambda_1 - \lambda_2 - \lambda_3 - \lambda_4\},
\end{eqnarray}
where \(\lambda_i\) are the square roots of the eigenvalues of the matrix \((\rho\tilde{\rho})^{1/2}\) in decreasing order, and \(\tilde{\rho}\) is the spin-flipped state of \(\rho\) \cite{Coffman2000Concurr}. Finally, the entanglement of formation (EoF) \cite{Bennett1996EoF, Wootters1998EoF} for a bipartite pure state \(\ket{\psi}_{AB}\) is determined by the entropy of a subsystem, specifically given by \(E_f(\ket{\psi}_{AB}) = S(\rho_A)\), where \(\rho_A = \text{tr}_B(|\psi\rangle{_{AB}}\langle\psi|)\) represents the reduced density matrix of \(\ket{\psi}_{AB}\) on subsystem \(A\). More generally, the von Neumann entropy \(S(\rho) = -\text{tr}(\rho \log \rho)\) is employed to quantify the information content of the quantum state \(\rho\). For a bipartite mixed state \(\rho_{AB}\), the EoF is defined as the minimum average entanglement over all possible pure state decompositions:
\begin{eqnarray}\label{EntofForm}
E_f(\rho_{AB}) = \min \sum_{i} p_i E_f(|\psi_i\rangle_{AB}),
\end{eqnarray}
where \(\rho_{AB} = \sum_i p_i|\psi_i\rangle_{AB}\langle\psi_i|\) represents the convex combination of pure states \(|\psi_i\rangle\) with associated probabilities \(p_i\). Entanglement measures have operational interpretations, meaning that they can be related to specific tasks or capabilities in QIP. For example, the entanglement of formation quantifies the amount of entanglement required to create a given entangled state. It represents the minimum amount of entanglement needed to prepare a particular state using only LOCC. The entanglement of formation is not directly related to the efficiency of entanglement distillation protocols, which aim to extract highly entangled states from partially entangled ones. Entanglement entropy, on the other hand, characterizes the amount of entanglement in a quantum state. It provides insights into the number of maximally entangled states that can be generated from a given state by LOCC. However, it is not directly linked to the efficiency of entanglement distillation protocols \cite{Plenio2007-QEntMeasure, Horodecki2009-QEnt}.

Section \ref{Sec:Majorization-Based} presents a brief introduction to majorization, which establishes the fundamental framework for subsequent discussions and analyses. The significance of majorization theory in the quantification and comparative analysis of entanglement levels in bipartite quantum states cannot be overstated \cite{Nielsen1999MAJ, NielsenVidal2001Majorization, Nielsen2002MAJIntroduction}. Nielsen's study \cite{Nielsen1999MAJ} holds a paramount position as it introduced and catalyzed subsequent investigations into the practical implementation and theoretical implications of this fundamental theory. Soon after, in the review by Nielsen and Vidal \cite{NielsenVidal2001Majorization}, key aspects of majorization theory were introduced and significant advancements in the field of bipartite entanglement transformations were explored. The majorization condition states that a quantum state, represented in Schmidt form as \(\ket{\psi}_{AB} = \sum_{i=0}^{d-1} \psi_i \ket{i}_A\ket{i}_B\), can be transformed into another state, denoted as \(\ket{\phi}_{AB} = \sum_{i=0}^{d-1} \phi_i \ket{i}_A\ket{i}_B\), via LOCC with unit probability if and only if the eigenvalues vector of the reduced density matrix, denoted as \(\lambda(\psi)\), is majorized by the eigenvalues vector of \(\lambda(\phi)\). This majorization relationship is expressed as \(\lambda(\psi) \prec \lambda(\phi)\), where \(\lambda(\psi)\) and \(\lambda(\phi)\) represent the decreasingly ordered vectors of eigenvalues obtained from the reduced density matrices \(\rho_{A}^{\psi}\) and \(\rho_{A}^{\phi}\), respectively. An example will now be provided to further facilitate the discussion. The maximally entangled state for the bipartite case in a \(d\)-dimensional system is represented as:
\begin{eqnarray}\label{MaxEntBipartite}
\ket{\Phi_{d}} = \frac{1}{\sqrt{d}} \sum_{i=0}^{d-1} \ket{i}_A\ket{i}_B.
\end{eqnarray}
The state \(\ket{\Phi_{d}}\) is considered maximally entangled (in Fig.~\ref{FIG:ElementsofQRTs}, for instance, we have \(\rho_{\star}:= \ket{\Phi_{d}}\bra{\Phi_{d}}\) for bipartite entanglement) as it achieves the highest possible degree of entanglement between the two subsystems. The state \(\ket{\Phi_{d}}\) given in Eq.~\eqref{MaxEntBipartite} can be effectively and deterministically transformed into any arbitrary state by means of LOCC with a probability of unity, i.e., \(\ket{\Phi_{d}}\) is majorized by any other state \(\ket{\psi}_{AB}\), \(\lambda(\Phi_{d}) \prec \lambda(\psi)\), where \(\lambda(\Phi_{d})= (\frac{1}{d}, \frac{1}{d}, \dots, \frac{1}{d})\). In closing, the influence of majorization extends to the domain of entanglement theory and warrants thorough exploration and discussion in appropriate contexts, as will become evident through the subsequent sections.

The field of quantum entanglement has undergone significant advancements since its inception, making it a pioneering area of research. Initially, exploratory investigations paved the way for a deeper understanding of this phenomenon. As time progressed, the studies in this field intensified, gaining momentum and resulting in a vast and substantial body of literature \cite{Horodecki2009-QEnt}. Today, the scientific community is equipped with a comprehensive and extensive knowledge base that encapsulates the intricacies of quantum entanglement. The evolution of research in quantum entanglement has been marked by remarkable growth and profound developments. The exploration of this intricate phenomenon has captivated the scientific community, prompting rigorous investigations and stimulating intellectual discourse. Significantly, entanglement has served as a paradigmatic example showcasing the utility of quantum-specific properties as valuable resources. Its profound implications have not only stimulated extensive research on entanglement itself but have also fostered the exploration and investigation of other fundamental quantum mechanical properties.


\subsection{Resource Theory of Coherence}\label{Sec:RTCoherence}

Quantum coherence pertains to the intrinsic property of quantum systems, wherein they exhibit a simultaneous existence in multiple states, enabling the occurrence of interference phenomena. It plays a pivotal role as a resource, enabling the implementation of quantum algorithms and facilitating computational speedup. Consequently, it is an actively researched phenomenon in emerging fields such as quantum metrology \cite{Giovannetti2006QMetrology, Toth2014QMetrology, Friis2017QMetrology, Shlyakhov2018QMetrology} and quantum algorithms \cite{Hillery2016CoherenceAsResource}. Within the scope of recent scientific inquiry, Streltsov, Adesso, and Plenio \cite{Streltsov2017-RTofCoh} explored and assessed the progress of this rapidly expanding field of research, which encompasses the study of quantum coherence in terms of its characterization \cite{Streltsov2017StructureQC, Carmeli2018CharacterizationQC, Mandal2020CharacterizingQC, Yamasaki2021hierarchyofquantum}, quantification \cite{Girolami2014QuantifyQC, Shao2015QuantifyingQC, Yuan2015QuantifyingQC, Streltsov2015QuantifyingQCohEnt, Yu2016QuantifyingQCAlter, Liu2017CohMeasure, Yu2017QuantifyingQC, Zhang2018QuantifyingQC, Dong2019DetectCoh, Ding2021QuantifyingQC, Designolle2021SetCohQuantify, Ding2021EffEstCoh, Bischof2021QuantifyingQC, Fu2022QuantifyingQC, Sun2022QuantifyingQC}, manipulation \cite{Du2015ConditionMajCoh, Winter2016OpRTofC, Torun2018DetCohTr, Chen2019OneShotCohDistill, Liu2019DetCohDistill, Torun2019DistQCoh, Lami2019ManipulationQCIOs, Lami2020ManipulationQC, Zhang2020OneShotQCDistill, Liu2020CatalystQCDitill, Regula2020QCManipDephasing, Yang2021AvarageQCDistill, Liu2021OptimalQCDistill, Cunden2021GenericQC, Ding_2022NoGo, Liu2022ApproxQCTr, Gour2022RoleofQC}, dynamical evolution \cite{Bromley2015FrozenQC, Huang2017DynamicsQC, Qin2018DynamicsQC, Radhakrishnan2019DynamicsQC, Wang2019DynamicsQC, Radhakrishnan2019DQC, Jafari2020DynamicsQC, Yin2022DynamcisQC}, and practical applications\cite{Masini2021CohInterferometry, Regula2021ExpProQCoh}. Therefore, to gain a more thorough grasp of the subject matter, we highly recommend consulting Ref.~\cite{Streltsov2017-RTofCoh}, as it encompasses a broader range of information surpassing the discussed scope here.

Coherence is a concept that relies on the choice of a specific basis. It is of utmost significance to emphasize that the selection of the reference basis should be aligned with the fundamental principles and laws of physics relevant to the problem at hand. By aligning the reference basis with the specific physics governing the system under consideration, one can ensure a comprehensive understanding and analysis of the coherence phenomena within the given context. In that connection, we consider a particular basis \(\{\ket{i}: i = 0, 1, \dots, d-1\}\) in the \(d\)-dimensional Hilbert space \(\mathcal{H}_d\), which we designate as our preferred basis. This basis is characterized by its completeness and orthonormality.  Quantum states that possess a diagonal representation in terms of this specific basis are referred to as incoherent states, constituting a set labeled by \(\mathcal{I}\) \cite{Aberg2006Quantifying, Baumgratz2014QCoherence}. Hence, all incoherent states \(\delta \in \mathcal{I}\) are of the form
\begin{eqnarray}\label{IncoherentStates}
\delta = \sum_{i=0}^{d-1} \delta_{i} \ket{i}\bra{i},
\end{eqnarray}
where \( \sum_{i} \delta_{i} = 1\) such that \(\delta_{i} \geq 0\) for all \(i = 0, 1, \dots, d-1\). The set of incoherent states forms a fundamental reference point for comparing and quantifying coherence \cite{Baumgratz2014QCoherence}. Coherence measures are defined based on the distinguishability between a given state and the set of incoherent states (again, see Fig.~\ref{FIG:ElementsofQRTs}). These measures provide a quantitative way to evaluate the coherence content of a state and track its transformation under various operations \cite{Du2015ConditionMajCoh, Winter2016OpRTofC, Torun2018DetCohTr, Chen2019OneShotCohDistill, Liu2019DetCohDistill, Torun2019DistQCoh, Lami2019ManipulationQCIOs, Lami2020ManipulationQC, Zhang2020OneShotQCDistill, Liu2020CatalystQCDitill, Regula2020QCManipDephasing, Yang2021AvarageQCDistill, Liu2021OptimalQCDistill, Cunden2021GenericQC, Liu2022ApproxQCTr, Gour2022RoleofQC}.

Similar to other resource theories, states that lie outside the set of free states are regarded as resources within the RTC, specifically referring to coherent states. A finite \(d\)-dimensional pure coherent state is given by
\begin{eqnarray}\label{ResourcePureCoherence}
|{\psi}\rangle = \sum_{i=0}^{d-1} e^{i\theta_i}\psi_i |{i}\rangle \quad \Big(\psi_i \in \mathds{R}, \; 0 \leq \theta_i \leq \pi\Big),
\end{eqnarray}
where \(\{\psi_i: i=0, 1, \dots, d-1\}\) are non-negative real numbers arranged in non-increasing order (\(\psi_k \geq \psi_{k+1} \geq 0\)), and satisfying \(\sum_{i=0}^{d-1} \psi_{i}^{2} = 1\). Needless to say, in order to facilitate an examination of quantum coherence, it is essential to consider Eq.~\eqref{ResourcePureCoherence} with multiple non-zero values of \(\psi_i\). This requirement arises from the intrinsic nature of coherence, which involves the presence of distinct quantum superposition states characterized by non-zero coefficients. By incorporating multiple non-zero values of \(\psi_i\) in Eq.~\eqref{ResourcePureCoherence}, we ensure the inclusion of the necessary components to explore and analyze the intricate phenomena associated with quantum coherence. Here, without loss of generality, we can and from now on will assume that \(\theta_i = 0\) for all \(i\), as all these complex phases can also be eliminated by diagonal unitaries, which are always assumed to be free operations in any version of the RTC. A compact mathematical summary for quantum states in RTC is then as follows:
\begin{eqnarray}
\left\{{\mathrm{C}_{\mathcal{F}} = \delta \in \mathcal{D}(\mathcal{H}_d) : \;  \delta =  \sum_{i=0}^{d-1} \delta_i \ketbra{i}{i}, \;  \delta_i \geq 0, \;  \sum_{i=0}^{d-1} \delta_i = 1}\right\},
\end{eqnarray}
\begin{eqnarray}
\Big\{\mathrm{C}_{\mathcal{R}} = \sigma \in \mathcal{D}(\mathcal{H}_d) : \; \sigma \notin \mathrm{C}_{\mathcal{F}}\Big\},
\end{eqnarray}
where \(\mathcal{D}(\mathcal{H}_d)\) denotes the set of density operators on \(\mathcal{H}_d\). Here, we use the notations \(\mathrm{C}_{\mathcal{F}}\) and \(\mathrm{C}_{\mathcal{R}}\) to refer to the sets of incoherent states and coherent states, respectively. Moreover, we denote by \(\ket{\Psi_d}\) the maximally coherent state in the reference basis with entries \(\psi_i=\frac{1}{\sqrt{d}}\),
that is,
\begin{eqnarray}\label{MaximallyCoherentState}
\ket{\Psi_d}=\frac{1}{\sqrt{d}}\sum_{i=0}^{d-1} e^{i\theta_i} \ket{i}.
\end{eqnarray}
Since all other \(d\)-dimensional coherent states can be generated from \(\ket{\Psi_d}\) by means of the free operations, this definition regards as a unit of RTC --- coherence bit \cite{Streltsov2017-RTofCoh}. Maximally coherent states hold a pivotal position in the realm of quantum information science and its associated disciplines. These states embody the utmost level of coherence attainable within a specific quantum system (in Fig.~\ref{FIG:ElementsofQRTs}, for instance, we have \(\rho_{\star} := \ket{\Psi_{d}}\bra{\Psi_{d}}\) for RTC), rendering them valuable resources for diverse quantum protocols and tasks \cite{Giovannetti2006QMetrology}. Comprehending and harnessing the potential of maximally coherent states are indispensable for unlocking the complete range of capabilities offered by quantum technologies and propelling the boundaries of quantum information science.

To delve into the dynamical characteristics of quantum coherence, a complete understanding of its evolution under suitable transformations, referred to as free operations, is imperative \cite{Streltsov2017-RTofCoh}. Ideally, one aims to ascertain these operations based on the inherent physical properties of the underlying resource. One possible strategy is to define free operations through a set of axiomatic considerations. This involves discerning the natural constraints that a specific class of free operations should meet and exploring the collection of all channels that satisfy these constraints. The fundamental premise guiding this approach is that free operations should not introduce any additional coherence. Within the scope of this study, we direct our attention towards several renowned operations (see Fig.~2 in Ref.~\cite{Chitambar2016IncohOperations}). These operations are widely recognized in the relevant literature \cite{Streltsov2017-RTofCoh}.

{\bf\emph{Incoherent operations}} --- The class of quantum operations known as incoherent operations (IOs) can be characterized by the presence of Kraus operators \(\{K_n\}\) that satisfy two key conditions. First, the operators must adhere to the condition \(\sum_{n} K_n^\dagger K_n = \mathds{1}\), ensuring the preservation of probability. Second, for any input state \(\delta\) belonging to the set of incoherent states \(\mathcal{I}\) (that is, \(\delta \in \mathcal{I}\)), the resulting output state \(\rho_n\) is given by
\begin{equation}
\rho_n = \frac{K_n \delta K_n^{\dagger}}{\mathrm{Tr}\left(K_n {\delta} K_n^\dagger\right)} \subseteq \mathcal{I}
\end{equation}
for all \(n\), where \(\mathrm{Tr}(\cdot)\) represents the trace operation. This implies that incoherent states remain incoherent under IOs. This property arises from the requirement that the Kraus operators \(\{K_n\}\) satisfy \(K_n \sigma K_n^{\dagger} \subset \mathcal{I}\) for all \(n\), where \(\sigma\) denotes any incoherent state. Consequently, any initial incoherent state subjected to an IO will result in a final state that remains within the set of incoherent states, preserving its incoherent nature throughout the operation.


{\bf\emph{Physically incoherent operations}} --- A coherence-free operation is referred to as a physically incoherent operation (PIO) if and only if it can be expressed as a convex combination of maps, where each map is uniquely defined by its corresponding set of Kraus operators such that
\begin{equation}
K_m = {U_m} {P_m} = \sum_{x} e^{i{\theta}_x} |{\pi_m(x)}\rangle \langle {x}| {P_m}.
\end{equation}
In the given context, a set of operators \(\{P_m: m = 1, 2, \dots, n\}\) is defined as an orthogonal and complete collection of incoherent projectors on the primary system. The function \({\pi}_m(x)\) represents permutations, which indicate the rearrangement or reordering of elements within the parameter \(x\).

{\bf\emph{Strictly incoherent operations}} ---  A strictly incoherent operation (SIO), denoted as $\mathcal{E}$, is characterized by a set of Kraus operators $K_j$ for $j=1, 2, \dots, n$. These operators possess a crucial property: For any input state $\rho$, the SIO transforms the dephased state $\Delta(\rho)$ in the same manner as the application of the dephasing operation to the transformed state $\Delta({K_j} \rho {K_j}^{\dagger})$ for all $j$; that is, \({K_j} {\Delta(\rho)}{K_j}^{\dagger} = \Delta({K_j} \rho {K_j}^{\dagger})\) for all \(j\). Here, the completely dephasing map \(\Delta(\cdot)\) is defined as follows:
\begin{equation}
\rho \mapsto \Delta(\rho) = \sum_{i=0}^{d-1} |{i}\rangle \langle {i}|\rho|{i}\rangle \langle {i}|.
\end{equation}
where \(\{|i\rangle: i=0, 1, \dots, d-1\}\) represents the basis states and \(d\) denotes the dimension of the system. This property of a SIO ensures the preservation of the dephased structure: The resulting state after the SIO transformation remains in a dephased form.

{\bf\emph{Dephasing-covariant incoherent operations}} --- A quantum operation $\mathcal{E}$ is considered dephasing-covariant with respect to the preferred subspaces if it commutes with the associated dephasing operation \(\Delta(\cdot)\). This condition is expressed as \(\mathcal{E} \circ \Delta = \Delta \circ \mathcal{E}\). An important consequence of dephasing-covariance is observed when considering incoherent states. Specifically, if \(\mathcal{E}\) is dephasing-covariant, it follows that for any incoherent state \(\delta \in \mathcal{I}\), the action of \(\mathcal{E}\) on \(\delta\) is equivalent to applying the dephasing operation first and then applying \(\mathcal{E}\). Mathematically, this can be expressed as \(\mathcal{E}(\delta) = \mathcal{E}(\Delta(\delta)) = \Delta(\mathcal{E}(\delta))\). Consequently, the resulting state \(\mathcal{E}(\delta)\) remains invariant under dephasing, which implies that it retains its incoherent nature. Thus, dephasing-covariant incoherent operations (DIOs) preserve the incoherence of states.

{\bf\emph{Maximal incoherent operations}} --- An operation \(\mathcal{E}\) is classified as a maximal incoherent operation (MIO) if it maps any incoherent state \(\delta\) given in Eq.~\eqref{IncoherentStates} to another incoherent state, i.e., \(\mathcal{E}(\delta) \in \mathcal{I}\) for all \(\delta \in \mathcal{I}\). The concept of MIOs is important in the study of quantum coherence. By exclusively affecting incoherent states while leaving coherent aspects untouched, MIOs provide a useful framework for understanding and controlling coherence in quantum systems. These operations play a crucial role in coherence theory, enabling the development of coherence-preserving quantum devices and coherent control strategies.

Coherence measures play a crucial role in characterizing and quantifying the degree of coherence present in quantum systems \cite{Girolami2014QuantifyQC, Shao2015QuantifyingQC, Yuan2015QuantifyingQC, Streltsov2015QuantifyingQCohEnt, Yu2016QuantifyingQCAlter, Yu2017QuantifyingQC, Zhang2018QuantifyingQC, Ding2021QuantifyingQC, Designolle2021SetCohQuantify, Bischof2021QuantifyingQC, Fu2022QuantifyingQC, Sun2022QuantifyingQC}. They provide a quantitative measure to assess the resourcefulness of coherent states for various QIP tasks. Two well-known proper measures of coherence are the relative entropy of coherence and the \(\ell_1\)-norm of coherence. The relative entropy of coherence, denoted as \(C_{\text{rel. ent.}}\), is given by:
\begin{equation}\label{RelEntofCoh}
C_{\text{rel. ent.}} = S(\rho_{\text{diag}}) - S(\rho),
\end{equation}
where \(S\) represents the von Neumann entropy and \(\rho_{\text{diag}}\) is the state obtained from \(\rho\) by deleting all off-diagonal elements. It quantifies the difference between the entropy of the original state and the entropy of the state with eliminated coherences. The \(\ell_1\)-norm of coherence, denoted as \(C_{\ell_1}\), is defined as:
\begin{equation}\label{L1Norm}
C_{\ell_1} = \sum_{i \neq j} |\rho_{ij}|,
\end{equation}
where \(\rho_{ij}\) denotes the off-diagonal elements of the state \(\rho=\sum_{i,j}\rho_{ij}\ket{i}\bra{j}\). This measure computes the sum of the absolute values of all off-diagonal elements, providing a measure of the total coherence in the state. For incoherent states described by Eq.~\eqref{IncoherentStates}, both coherence measures yield zero, i.e., \(C_{\text{rel.ent}}(\delta) = C_{\ell_1}(\delta) = 0\). This result confirms that incoherent states have no off-diagonal elements and, therefore, lack coherence according to these measures.

Moreover, the robustness of coherence, initially proposed by Napoli \emph{et al}. \cite{Napoli2016RobustCoh} and Piani \emph{et al}. \cite{Piani2016RobustCoh}, serves as an essential coherence monotone. It quantifies the minimum amount of mixing necessary to render a given state \(\rho\) incoherent. This measure, denoted as \(R_C(\rho)\), is computed by optimizing over all quantum states \(\tau\) and is defined as:
\begin{equation}\label{RobustofCoh}
R_C(\rho) = \min_{\tau} \left\{s \geq 0 {\bigg|} \frac{\rho + s\tau}{1 + s} \in \mathcal{I}\right\},
\end{equation}
where the parameter \(s \geq 0\) ensures the mixing preserves the incoherent set \(\mathcal{I}\). For single-qubit states, \(X\) states, and pure states, the robustness of coherence \eqref{RobustofCoh} coincides with the well-known \(\ell_1\)-norm of coherence given by Eq.~\eqref{L1Norm}. Further insights and comprehensive discussions on coherence measures can be found in Ref.~\cite{Streltsov2017-RTofCoh}, offering readers a detailed exploration of various quantifiers, their properties, and their applications in quantum information science.

Now, we turn our attention to the examination of majorization within the context of RTC. This examination aligns with the discourse presented in Sec.~\ref{Sec:Majorization-Based}, particularly concerning its implications and relevance in the domain of quantum entanglement. The criterion of majorization serves as a reliable indicator of the feasibility of transforming one coherent state into another under IOs. Drawing motivation from the seminal work by Nielsen \cite{Nielsen1999MAJ}, Du \emph{et al.} \cite{Du2015ConditionMajCoh} successfully demonstrated that a coherent state \(\ket{\psi}=\sum_{i=0}^{d-1} \psi_i \ket{i}\) with \(\psi_k \geq \psi_{k+1}> 0\) can be deterministically transformed into another coherent state \(\ket{\phi}=\sum_{i=0}^{d-1} \phi_i \ket{i}\) with \(\phi_k \geq \phi_{k+1}> 0\) (for \(k=0, 1, \dots, d-2\)). This transformation is feasible if and only if the coherence vector \(\mu{(\psi)}=(\psi_0^2, \psi_1^2, \dots, \psi_{d-1}^2)^{\mathrm{T}}\), as defined in \cite{Zhu2017CoherenceVector}, is majorized by the coherence vector \(\mu{(\phi)}=(\phi_0^2, \phi_1^2, \dots, \phi_{d-1}^2)^{\mathrm{T}}\), denoted by \(\mu{(\psi)} \prec \mu{(\phi)}\). Simply, the majorization condition is satisfied when the following inequalities hold for any \(k\in [0,d-2]\):
\(\sum_{i=0}^{k}\psi_i^{2} \leq \sum_{i=0}^{k}\phi_i^{2}\),
with the additional requirement of equality when \(k=d-1\) \cite{Du2015ConditionMajCoh}. These inequalities provide a quantitative measure of the majorization relationship between the coherence vectors \(\mu{(\psi)}\) and \(\mu{(\phi)}\), reflecting the extent to which one coherence vector dominates the other. The detailed analysis of deterministic transformations involving coherent states, with a particular emphasis on the concept of majorization, can be found in Ref.~\cite{Torun2018DetCohTr}.

Emphasizing the orthonormality of basis states, coherence plays a pivotal role in quantum advantage and enhancing QTs. Ongoing research aims to understand coherence, develop measures for manipulation, and explore its operational significance. Future directions involve preserving and controlling coherence, advancing coherence-based protocols, and uncovering foundational aspects. Besides all this, investigating the effects of nonorthogonal basis states is important. The presence of nonorthogonal basis states necessitates the exploration of a novel resource theory in this context, as addressed in Sec.~\ref{Sec:RTSuperposition}.


\subsection{Resource Theory of Superposition}\label{Sec:RTSuperposition}

The concepts of coherence and superposition share a conceptual similarity, but they exhibit distinct characteristics in their resource-theoretic formulations. A crucial distinction arises in the treatment of basis states. In the case of RTC \cite{Baumgratz2014QCoherence}, as we have discussed in Sec.~\ref{Sec:RTCoherence}, basis states are defined to be orthogonal to each other. However, in the resource theory of superposition (RTS) \cite{Theurer2017RTS}, basis states are not necessarily orthogonal, thereby introducing a significant departure from the RTC framework.

In the case of RTS, as introduced in Ref.~\cite{Theurer2017RTS}, the set of free states is defined as the set of density operators that do not exhibit any superposition. We use the notation \(\mathrm{S}_{\mathcal{F}}\) to refer to this specific set. Specifically, let \(\mathcal{H}_d\) be a \(d\)-dimensional Hilbert space, and let \(\{\ket{c_k}: k = 0, 1, \dots, d-1\}\) be a set of nonorthogonal (more generally, complete but not necessarily orthogonal), normalized, and linearly independent basis vectors of $\mathcal{H}_d$. Then, states defined as
\begin{eqnarray}\label{Superposition-Free-States}
\varrho =  \sum_{k=0}^{d-1} \varrho_k \ketbra{c_k}{c_k},
\end{eqnarray}
are called superposition-free, where \(\varrho_k \geq 0\) form a probability distribution, that is, \(\sum_{k}\varrho_k = 1\). All density operators which are not an element of \(\mathrm{S}_{\mathcal{F}}\) --- the set of superposition-free density operators --- are called superposition states and form the set of resource states. We use the notation \(\mathrm{S}_{\mathcal{R}}\) to refer to this specific set. On the other hand, a linear combination of \(\{\ket{c_k}: k = 0, 1, \dots, d-1\}\) gives us the pure superposition states
\begin{eqnarray}
\ket{\psi} = \sum_{k=0}^{d-1} \psi_{k} \ket{c_k},
\end{eqnarray}
where the coefficients \(\{\psi_k: k = 0, 1, \dots, d-1\}\) are complex numbers. Let \(G\) be overlap matrix of basis states\(\{\ket{c_k}\}\), i.e., Gram matrix, with the elements \(G_{ij}=\langle{c_i}|{c_j}\rangle\). Then the normalization condition reads \(\langle{\psi} |{\psi}\rangle=\sum_{i,j=0}^{d-1} \psi_{i}^{\ast} G_{ij} \psi_{j}=1\), where \(G_{ij}\) are complex in general. A compact mathematical summary for quantum states in RTS is then as follows:
\begin{eqnarray}
\left\{{\mathrm{S}_{\mathcal{F}} = \varrho \in \mathcal{D}(\mathcal{H}_d) : \;  \varrho =  \sum_{k=0}^{d-1} \varrho_k \ketbra{c_k}{c_k}, \;  \varrho_k \geq 0, \;  \sum_{k=0}^{d-1} \varrho_k = 1}\right\},
\end{eqnarray}
\begin{eqnarray}
\Big\{\mathrm{S}_{\mathcal{R}} = \varsigma \in \mathcal{D}(\mathcal{H}_d) : \; \varsigma \notin \mathrm{S}_{\mathcal{F}}\Big\},
\end{eqnarray}
where \(\mathcal{D}(\mathcal{H}_d)\) denotes the set of density operators on \(\mathcal{H}_d\).

\begin{table}[!t]
\centering
\begin{tabular}{| c | c | c |} \toprule \toprule
{Ingredients} & {Coherence} & {Superposition}
\\ \midrule \toprule
& \(\{\ket{i} : i=0, 1, \dots, d-1\}\)  &  \(\{\ket{c_k} : k=0, 1, \dots, d-1\}\)   \\
Basis States    & {reference basis;} & {normalized and linear independent;}   \\
& {complete and orthonormal} & {complete, but not necessarily orthogonal}   \\ \midrule \toprule
& \(\delta=\sum_{i} \delta_i \ket{i}\bra{i} \in \mathrm{C}_{\mathcal{F}}\) &  \(\varrho = \sum_k \varrho_k \ketbra{c_k}{c_k} \in \mathrm{S}_{\mathcal{F}}\) \\
Free States (\(\mathcal{F}\))    &   &    \\
& \(\delta_i \in [0,1]\) such that \(\sum_{i} \delta_i = 1\) &  \(\varrho_k \geq 0\), probability distribution \\ \midrule \toprule   &    &    \\
Resource States (\(\mathcal{R}\))   & any state \(\sigma \notin \mathrm{C}_{\mathcal{F}}\) (i.e., \(\sigma \in \mathrm{C}_{\mathcal{R}}\)) &  any state \(\varsigma \notin \mathrm{S}_{\mathcal{F}}\) (i.e., \(\varsigma \in \mathrm{S}_{\mathcal{R}}\)) \\    &    &   \\ \midrule \toprule
  & incoherent operation (IO); &  superposition-free  \\
Free Operations (\(\mathcal{O}\))  &   &  \\
& MIO, DIO, IO, SIO, PIO  &  \Big(\(K_n \varrho K_n^{\dagger} \in \mathcal{F}\) for all \(\varrho \in \mathcal{F}\)\Big)   \\ \bottomrule \bottomrule
\end{tabular}
\caption{The comparison of the resource-theoretic formulations of coherence and superposition. Emphasizing their contrasting characteristics, coherence is defined by the requirement of orthogonal basis states within its resource theory, while the resource theory of superposition allows for the inclusion of non-orthogonal basis states. By juxtaposing the essential elements of both theories, this table summarizes the nuanced differences.}
\label{Table:Ingredients}
\end{table}

Regarding the free operations within the RTS, the set of free operations, denoted by \(\mathcal{O}\), consists of all quantum operations that can be applied to the free states without creating superposition. Specifically, a Kraus operator \(K_n\) is called superposition-free if \(K_n \varrho K_n^\dagger \in \mathrm{S}_{\mathcal{F}}\) for all \(\varrho \in \mathrm{S}_{\mathcal{F}}\), and is of the form \cite{Theurer2017RTS}:
\begin{equation}
K_n = \sum_{k} c_{k,n} {\ket{c_{f_{n}(k)}}\bra{c_k^{\perp}}},
\end{equation}
where \(c_{k,n}\) are complex numbers, \({f_{n}(k)}\) are arbitrary index functions, and \(\langle{c_i^{\perp}}|{c_j}\rangle=\zeta_i\delta_{ij}\) for \(\zeta_i\in \mathds{C}\), and the vectors \(\ket{c_k^{\perp}}\) are normalized. Moreover, a quantum operation \(\Phi(\cdot)\) is called superposition-free if it is trace-preserving and can be written such that \(\Phi(\rho) = \sum_{n} K_n \rho K_n^\dagger\), where all \(K_n\) are free. Upon contrasting RTS with RTC, it becomes evident that RTC encompasses a more extensive spectrum of free operations. This observation thereby mandates a more exhaustive examination of the realm of free operations within the context of RTS \cite{Theurer2017RTS}.

Table \ref{Table:Ingredients} provides a brief overview of the main elements in the research areas of RTC and RTS. It serves as a handy reference for summarizing and comparing the key aspects of these two branches in the field of QRT. However, RTS fails to consider the overlaps that arise due to the quantum state indistinguishability associated with the nonorthogonality of basis states. Pusuluk \cite{pusuluk2022unified} demonstrated that neglecting these overlaps can lead to conceptual inconsistencies, as they can give rise to quantum correlations. To illustrate this point, we can revisit the free states \(\varrho\) defined in Eq. \eqref{Superposition-Free-States} by considering a four-level system consisting of only two states \(|c_1\rangle\) and \(|c_2\rangle\) with a real-valued overlap \(\langle c_1|c_2\rangle = s \in \mathds{R}\). Let us focus on a specific partition of this system into two two-level subsystems \(A\) and \(B\), such that \(|c_i\rangle := |a_i\rangle_A \otimes |b_i\rangle_B\) with \(\langle a_1|a_2\rangle = \langle b_1|b_2\rangle = \sqrt{s}\). In this scenario, the state
\(\varrho\) is given by:
\begin{equation} \label{Eq::QCState}
\varrho = p \, |a_1\rangle \langle a_1 | \otimes |b_1\rangle \langle b_1 | + (1 - p) \, |a_2\rangle \langle a_2 | \otimes |b_2\rangle \langle b_2 | \, ,
\end{equation}
which exemplifies quantum-quantum states. Although it remains a superposition-free state in the basis \(\{|c_1\rangle, |c_2\rangle\}\), it possesses nonclassical correlations in the form of quantum discord~\cite{Vedral-2001, Zurek-2002, 2010_PRL_VV_CorrWithRelEnt}.

A comprehensive resource theory (RT) of quantum discord has not yet been established in the literature. However, we do know that incoherent operations cannot create discordant states like (\ref{Eq::QCState}) without depleting the local quantum coherences initially present in the subsystems~\cite{2016_PRL_Benjamin}. In other words, the superposition-free state under consideration maintains coherence in any orthogonal basis acquired through local unitary transformations. Therefore, the current definition of superposition in resource theory \cite{Theurer2017RTS} does not sufficiently encompass quantum correlations as a subset \footnote{One may consider the existence of a similar issue in RTC. In Eq.~\eqref{Eq::QCState}, if we assume that \(\langle a_1|a_2\rangle = 0\) and \(\langle b_1|b_2\rangle = s\), then the state \(\varrho\) becomes an incoherent state with right-discord (a classical-quantum state). However, to comprehensively investigate the relationship between coherence and discord, it becomes crucial to express the state in a local orthogonal basis as in Refs.~\cite{2015_PRA_CoherenceAndDiscord, 2016_PRA_CorrelatedCoherenceAndDiscordAndEnt, 2017_PRA_DiscordlikeCorrOfCoh, 2018_PhysicsReports_CorrelatedCoherenceAndGeoDiscord}, ensuring that \(\varrho\) consistently contains coherence.}.

Ref.~\cite{pusuluk2022unified} introduced a unified framework called \textit{genuine} superposition (GS) to address these inconsistencies. The framework encompasses quantum superposition and quantum state indistinguishability, serving as the fundamental concept of nonclassicality from which quantum coherence and correlations can arise. The mathematical representation of quantum states in resource theory of genuine superposition can be summarized as follows:
\begin{eqnarray}
\left\{{\mathrm{GS}_{\mathcal{F}} = \varrho \in \mathcal{D}(\mathcal{H}_d) : \;  \varrho =  \sum_{k=0}^{d-1} \varrho_k \ket{c_k}\bra{c_k^{\perp}}, \;  \varrho_k \geq 0, \;  \sum_{k=0}^{d-1} \varrho_k = 1}\right\},
\end{eqnarray}
\begin{eqnarray}
\Big\{\mathrm{GS}_{\mathcal{R}} = \varsigma \in \mathcal{D}(\mathcal{H}_d) : \; \varsigma \notin \mathrm{GS}_{\mathcal{F}}\Big\}.
\end{eqnarray}

According to the given framework, a general density operator residing in \(\mathcal{H}_d\) can be represented as follows:
\begin{equation} \label{Eq::State::BiO}
\rho = \sum_{i,j=0}^{d-1} \rho_{ij} \ket{c_i} \bra{c_j^\perp}.
\end{equation}
In this equation, the coefficients \(\rho_{ij} = \langle c^\perp_i | \hat{\rho} | c_j \rangle\) form a non-Hermitian but trace-one matrix. The eigenvalues of this matrix are real. The elements along the diagonal of the matrix indicate the relative weights of the basis states \(\{|c_i\rangle\}\) in the density operator \(\rho\). These diagonal elements correspond to the components of the vectors used to study majorization relations, which are employed to investigate transformations between pure states in RTS~\cite{2021_PRA_GT}.

In comparison to RTS, RTGS not only offers a more consistent framework for nonorthogonal systems but also proves to be more practically advantageous. The quantification and manipulation of nonclassicality in optical coherent states can be viewed as an application of RTS~\cite{Theurer2017RTS}. In the realm of molecular electronic states, the significance of overlaps between nonorthogonal states becomes more evident, as chemical bonding and related quantum chemical phenomena arise from these overlaps. Aromaticity, which is one such phenomenon, lacks a universally accepted definition. However, a recent proof-of-concept study~\cite{mahir2023aroma} demonstrated that the amount of genuine superposition shared between atomic orbitals effectively captures the aromaticity order of typical aromatic molecules and aligns with existing measures of aromaticity. Conversely, RTS is incapable of providing the aromaticity order of these molecules.

The presence of nonorthogonal basis states in the RTS gives rise to intriguing implications and challenges. Unlike in RTC, where orthogonal basis states form a convenient reference for coherence quantification and manipulation, the nonorthogonality of basis states in RTS requires the consideration of alternative approaches. The nonorthogonality of basis states necessitates the development of new mathematical formalisms and tools to capture and characterize superposition phenomena accurately. Coherence measures and coherence manipulation techniques devised in the context of coherence theory may not be directly applicable or meaningful within the realm of superposition theory due to the absence of orthogonality constraints. In light of this, it becomes imperative to investigate potential tools or methodologies that can effectively bridge the gap between coherence and superposition. The Löwdin symmetric orthogonalization method stands out with remarkable attributes in this regard. A succinct summary of this method can be outlined as follows:

{\bf\emph{L\"{o}wdin symmetric orthogonalization}} --- Recall that \(\{\ket{c_0},\ket{c_1},\dots,\ket{c_{d-1}}\}\) be a set of nonorthogonal, normalized, and linearly independent basis vectors of the \(d\)-dimensional Hilbert space \(\mathcal{H}_d\) \cite{Theurer2017RTS}. With the help of the method of L\"{o}wdin symmetric orthogonalization \cite{Lowdin1950, PIELA2014e99} (abbreviated by LSO), we obtain an orthonormal basis set \(\{\ket{l_0},\ket{l_1},\dots,\ket{l_{d-1}}\}\) called as ``L\"{o}wdin basis'' \cite{Lowdin1950}. Once again, it is important to note that the RTC is inherently dependent on the choice of basis, allowing for the selection of a reference basis aligned with the specific physics of the problem at hand. In this context, we here adopt the L\"{o}wdin basis as our preferred basis, characterized by its completeness and orthonormality.

Let \(\bm{X}=\{{x_0},\dots,{x_{d-1}}\}\), with the labels \(\bm{X}\in\{\bm{C},\bm{L}\}\) and \(x\in\{c, l\}\), represents a set of linearly independent vectors. We can define a general (linear) transformation \(\mathbb{T}\) such that \(\bm{L} = \mathbb{T} \bm{C}\). Note that here and throughout, the vector \(x_i\) is used as equivalent to \(\ket{x_i}\). The linear transformation \(\mathbb{T}\) is obtained from the overlap matrix \(G\), with {{(\(i,j\))-th} matrix element given by \(G_{ij}=\langle{c_i}|{c_j}\rangle\), satisfying \(\mathbb{T}\mathbb{T}^{\dagger}=G^{-1}\). Since \(G\) is a positive Hermitian matrix, and therefore, have positive eigenvalues \(\{\lambda_k\}\), it can always be unitarily diagonalized by a unitary matrix \(U\) such that \(G_{\text{diag}} = U^{\dagger} G U\), from which we can get \({G^{{1}/{2}}_{\text{diag}}}=\text{diag}(\sqrt{\lambda_0}, \dots, \sqrt{\lambda_{d-1}})\). By using this matrix, we define the matrix \({G^{{1}/{2}}}=U G^{{1}/{2}}_{\text{diag}}U^{\dagger}\) and get \({G^{-{1}/{2}}}={(G^{{1}/{2}})}^{-1}\). Then LSO given by \(\bm{L} = \mathbb{T} \bm{C}\) reads \((\bm{L})_{j} = \sum_{i} ({G^{-{1}/{2}}})_{ji} (\bm{C})_{i}\), where \(({G^{-{1}/{2}}})_{ji}\) is the {{(\(j,i\))-th} matrix element of \({G^{-{1}/{2}}}\). Thus, we have
\begin{eqnarray}\label{LSOforLiasCi}
\ket{l_j}=\sum_{i=0}^{d-1}\left({G^{-{1}/{2}}}\right)_{ji}\ket{c_i}, \ \ \text{with} \ \ j=0,1,\dots,d-1.
\end{eqnarray}
It is important to acknowledge that there exist infinitely many methods to transform a nonorthogonal set into an orthonormal set. However, the distinguishing characteristic of the LSO method lies in its ability to guarantee \(\sum_{i} \|c_i - l_i\|^2=\min\), where \(\|c_i - l_i\|^2 \equiv \langle{c_i - l_i}|{c_i - l_i}\rangle\) \cite{Lowdin1950}. This unique attribute ensures a smooth and minimal alteration from the nonorthogonal basis to the orthogonal basis, preserving the inherent structure of the original vectors.

\begin{figure}[t!]
	\centering
	\includegraphics[width=.64\columnwidth]{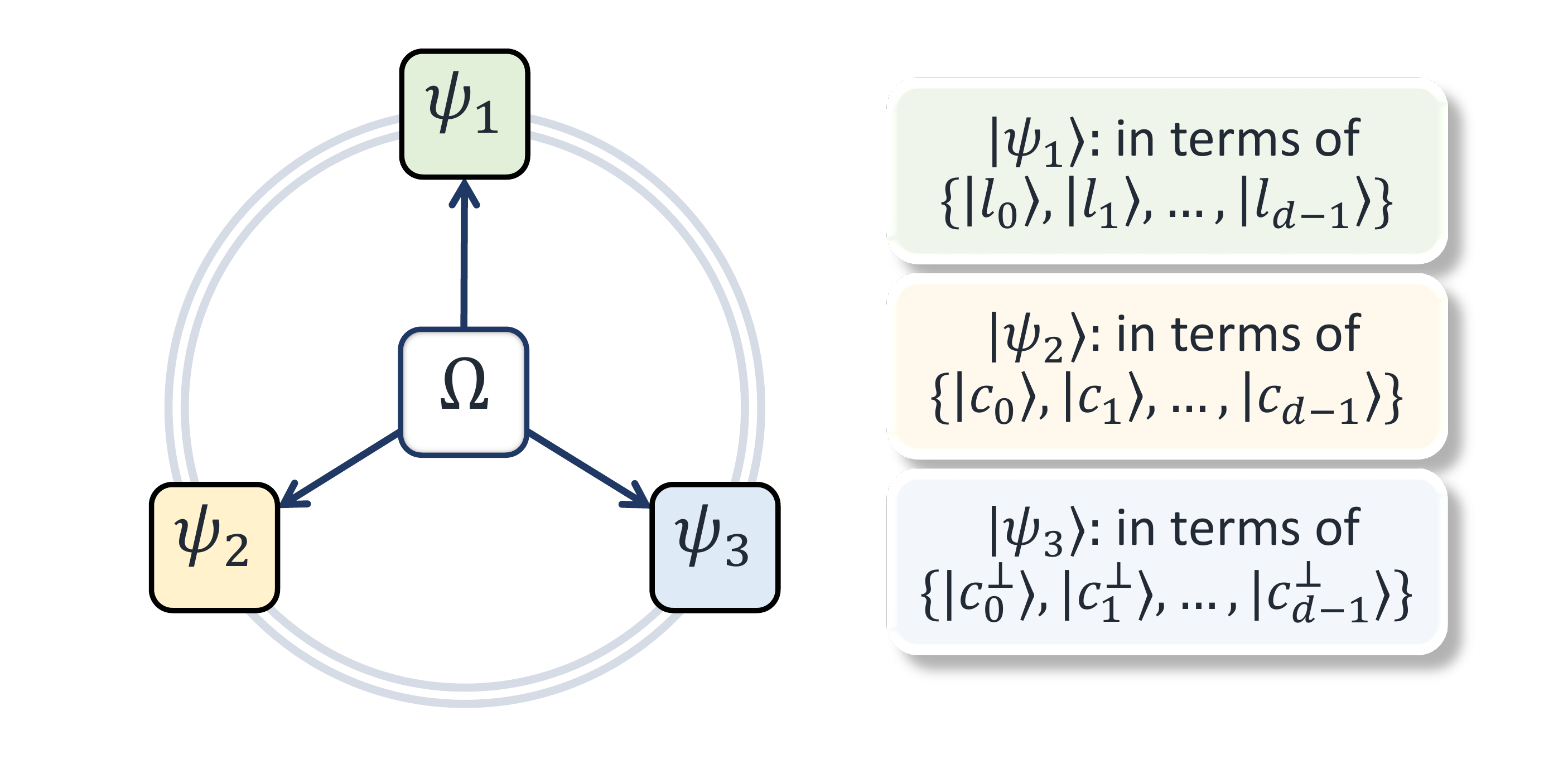}
	\caption{The Löwdin symmetric orthogonalization (LSO) method emerges as the approach that establishes a connection between quantum states \(\ket{\psi_1}\), \(\ket{\psi_2}\), and \(\ket{\psi_3}\), all representing the same quantum state \(\ket{\Omega}\) but with respect to different bases: \(\{\ket{l_{0}}, \ket{l_{1}}, \dots, \ket{l_{d-1}}\}\), \(\{\ket{c_{0}}, \ket{c_{1}}, \dots, \ket{c_{d-1}}\}\), and \(\{\ket{c_{0}^{\perp}}, \ket{c_{1}^{\perp}}, \dots, \ket{c_{d-1}^{\perp}}\}\), respectively. The method of LSO demonstrates the equivalence of maximally coherent states and states with maximal superposition, emphasizing the structural similarities between coherence and superposition (see Ref.~\cite{Torun2023Lowdin}). This highlights the analytical utility of LSO in analyzing resource states. The paper provides an illustrative example specifically focusing on two-dimensional cases.}
	\label{FIG:FIGCohSup00}
\end{figure}

Importantly, in Ref.~\cite{Torun2023Lowdin}, it was demonstrated that the states with maximal superposition, also known as golden states, can be obtained from the maximally coherent state using the LSO method. These findings align with the comprehensive study presented by Şenyaşa and Torun \cite{Senyasa2022Golden}, providing consistent results. As shown in \cite{Torun2023Lowdin}, the method of LSO yields a significant result:
\begin{eqnarray}\label{QubitMaxCoherentGen}
\ket{\Omega} & \equiv & \Big\{\ket{\psi_1}; \,  \ket{\psi_2};  \, \ket{\psi_3}\Big\} \nonumber \\
&=& \left\{
\frac{1}{\sqrt{2}}\left({\ket{l_0} \mp \ket{l_1}}\right); \,  \frac{1}{\sqrt{2(1 \mp {s})}}(\ket{c_0} \mp \ket{c_1});  \,  \frac{1}{\sqrt{2(1 \pm {s})}}(\ket{c_0^{\perp}} \mp \ket{c_1^{\perp}})\right\}.
\end{eqnarray}
The distinctive feature of the LSO method lies in its ability to establish a connection between maximally coherent states and states with maximal superposition. Notably, in the case of two-dimensional systems, maximally coherent states and states with maximal superposition are equivalent under the LSO method. This equivalence serves to emphasize the underlying structural similarities between coherence and superposition, demonstrating the analytical utility of LSO in the analysis of resource states within the context of the RTC and RTS (see Fig.~\ref{FIG:FIGCohSup00}).

Moreover, the nonorthogonal nature of basis states in the RTS leads to novel avenues for exploration. Researchers must devise alternative strategies to quantify and manipulate superposition resources effectively \cite{2021_PRA_GT}. This requires the careful analysis of the unique properties and mathematical structures arising from nonorthogonal basis states, paving the way for advancing our understanding of the RTS and its implications in various QIP tasks. Moreover, in-depth exploration and research in the field of RTS are crucial to establish a comprehensive understanding and facilitate meaningful comparisons with studies conducted within the realm of RTC. To bridge the knowledge gap and advance the understanding of quantum resource theories, additional research efforts are warranted to delve into the intricacies of RTS and elucidate its distinctive features in relation to RTC.


\section{Nonequilibrium Quantum Thermodynamics}\label{Sec:ThermodynamicsApproach}

Quantum coherence and correlations are essential resources arising from the delocalization of quantum information across space. Increasing attention is being given to understanding the thermodynamic methods used to create and protect these quantities \cite{2015_NewJPhys_Brunner_T2Ent, 2015_NJP_Brunner_TDCostOfCorrelations, 2017_PRA_Ozgur, 2018_QIP_EntIn2QubitsWith2CommonBaths, 2018_Quantum_BrunnerT2Ent, 2019_PRE_T2QCoh, 2020_PRA_Brunner_AutonomousMultipartiteEntEngines}. Moreover, they provide opportunities to manipulate the transfer of thermodynamic energy \cite{AHF_2008_PRE_Partovi, Lutz2009, AHF_2010_PRE_JenningsAndRudolph, 2016_Entropy_Ozgur, 2018_PRE_RoleOfQCohInHT, AHF_2019_QuantumSciTechnol_Petruccione_AppTs, AutoThermalMach_2019, 2019_npj, 2019_PRE_Ozgur_Multiatom, 2019_PRL_Esposito_TDofInfoFlow, 2019_PRA_CohNeeded4HF, AHF_2019_NatCommun_Lutz, 2019_OSID, AHF_2019_PhysRevResearch_Petruccione, 2019_PRL_Esposito_PiWithBathCoh,  2020_PRA_HorizantolCohAndPops, AHF_2020_arXiv_IonTraps, 2020_arXiv_2006_01166}. Furthermore, both the delocalized quantum information and heat flows can be mathematically described through quantum Onsager relations, indicating a reciprocal relationship between these two different kinds of resources~\cite{pusuluk2021prr}. However, the studies focusing on the transformations between quantum information and energy have mostly adopted an approach based on open quantum systems theory. Can we also investigate the same relationship using an approach based on RT? For this purpose, we need to extend the majorization-based RT framework to incorporate thermodynamics. In the following sections, we will outline how to accomplish this from simpler to more complex quantum systems far away from equilibrium.


\subsection{Resource Theory of Nonuniformity}\label{Sec:Nonuniformity}

To elucidate the duality between quantum information and disordered energy as resources~\cite{2017_NJP_ChiribellaAndScandalo_GPTs}, we can begin by revisiting the theory of  bipartite entanglement, which is limited by the transformation of pure states. In a pure composite system, bipartite entanglement is directly related to the local uncertainty in subsystems. Therefore, the entropy of one subsystem can be used as a measure of entanglement. However, the notion of entropy is insufficient to determine whether one entangled pure state can be transformed into another using LOCC. For this purpose, we need to use a stricter measure of disorder, which is the majorization criterion discussed in Sec.~\ref{Sec:Majorization-Based}.

The manifestation of majorization pre-order is not limited to the transformation law of bipartite entanglement facilitated by LOCC, as demonstrated in Refs.~\cite{2003_PRL_NO, 2003_PRA_NO}. There are other scenarios worth considering, such as deterministic transformations from less entangled pure states to more entangled pure states, which correspond to the inverse of LOCC transformations. In such cases, increasing entanglement requires a concurrent increase in the local uncertainty of the subsystems. This can only be achieved through a transformation that exhibits its impact on subsystems by means of a doubly-stochastic matrix. This is where majorization comes into play. If a probability distribution \(\vec{p}\) can be transformed into another probability distribution \(\vec{q}\) using a doubly-stochastic matrix \(B\), then \(\vec{p}\) must majorize \(\vec{q}\), i.e., \(B \, \vec{p} = \vec{q}\) iff \(\vec{p} \succ \vec{q}\) where \(\Sigma_i B_{ij} = \Sigma_j B_{ij} = 1\) (\(B_{ij} \geq 0\)). In other words, in order to increase the  entanglement shared between subsystems, the final spectrum of each reduced state should be majorized by its respective version before the transformation. LOCC assists the opposite of this transformation. Therefore, the probability distribution that describes the subsystems after the transformation should be majorized by its respective version before the transformation, i.e., \(\text{LOCC}: \rho_{AB} \mapsto \Lambda(\rho_{AB}) = \sigma_{AB}\) iff \(\vec{q} \prec \vec{p}\) such that \(\rho_{A/B} = diag(\vec{p})\) and \(\sigma_{A/B} = diag(\vec{q})\) .

Similarly, by working in the opposite scheme of bipartite entanglement theory, majorization characterizes the possible state transformations in the RT of nonuniformity (or equivalently, purity~\cite{2017_NJP_ChiribellaAndScandalo_GPTs}). In this framework, the operations that can be performed without cost are limited to those that introduce noise to the system or implement reversible changes. As athermality includes nonuniformity as a subset (see Fig.~\ref{FIG:Sum}), this scenario can be seen as a specific case where agents can perform a set of thermal operations at a given temperature. In such cases, all systems have trivial Hamiltonians, which is equivalent to considering the infinite temperature limit. Consequently, the only state that is free under \textit{noisy operation} (NOs)~\cite{2003_PRL_NO, 2003_PRA_NO, 2015_PhysReport_Nicole_Review} is the maximally mixed state:
\begin{equation}
\varrho = \frac{1}{d} \, \mathds{1},
\end{equation}
where \(d\) is the dimension of \(\mathcal{H}_d\) and \(\mathds{1}\) is the identity operator acting on this Hilbert space. This diagonal matrix corresponds to a uniform distribution of probabilities, reflecting an equilibrium state of information. That is to say, \(\varrho = diag(\vec{u})\) and \(\vec{u} = \{1/d\}_{j=1}^d\). Any other states are considered out-of-equilibrium, thereby making them valuable resource states within the context of NOs. The degree to which these states deviate from uniformity is commonly referred to as their nonuniformity, which serves as the primary resource of interest in this context.

Given a quantum system with a state \(\rho\) and a Hilbert space \(\mathcal{H}_d\), NOs are CPTP maps \(\mathcal{E}: \mathcal{L}(\mathcal{H}_d) \rightarrow \mathcal{L}(\mathcal{H}_d)\) of the form
\begin{equation}\label{Eq::NOs}
    \mathcal{E}(\rho) = \mathrm{tr}_{anc}\left[U (\rho \otimes \varrho_{anc}) U^\dagger\right],
\end{equation}
where \(\varrho_{anc} = \mathds{1}_{anc}/d_{anc}\) is the informational equilibrium state of an ancillary system and \(U\) is an arbitrary unitary operation acting on the joint system. We can interpret the operation described in Eq.~\eqref{Eq::NOs} using its decomposition into three steps as follows. An agent can couple the system of interest with an ancillary system in equilibrium. Then, it can allow any interaction between these two systems, effectively scrambling the information throughout the entire system while maintaining the total entropy intact. And finally, the agent can detach the ancillary system from the system of interest.

As mentioned earlier, a majorization-based investigation of LOCC-assisted bipartite entanglement transformations is only feasible when the initial and target states are pure. Similarly, the possibility of state transformations under NOs can only be examined through majorization when these states are mixed. If \(\rho = diag(\vec{p})\) transforms to \(\sigma  = diag(\vec{q})\) using NOs, it is necessary and sufficient that \(\vec{p} \succ \vec{q}\). These systems are referred to as isolated because they can solely exchange information with their surroundings as shown in Fig.~\ref{FIG:Sum}. In conventional thermodynamics, entropy is the thermodynamic quantity that appears in the inequality characterizing their state transformations. However, the majorization criterion mentioned above imposes a more stringent constraint than conventional thermodynamics by defining multiple inequalities based on the system's size. These constraints apply to thermodynamic state transformations of out-of-equilibrium, microscopic, and strongly correlated systems. As we approach the thermodynamic limit, all these constraints converge to a single entropic inequality that holds for uncorrelated equilibrium systems. This convergence occurring at the thermodynamic limit can be demonstrated through nonuniformity monotones.

The order-\(\alpha\) R\'{e}nyi divergences, given by the equation
\begin{eqnarray}\label{Eq::RenyiDiv}
    S_\alpha[\vec{x}\,||\,\vec{y}] = \frac{\text{sgn}(\alpha)}{\alpha -1} \log \sum_i x_i^\alpha y_i^{1- \alpha} ,
\end{eqnarray}
serve as a comprehensive set of nonuniformity monotones. Here, \(\alpha \geq 0\) and \(\alpha \neq 1\). When \(\vec{q} \prec \vec{p}\), it follows that \(S_\alpha[\vec{q}\,\,||\,\vec{u}] \leq S_\alpha[\vec{p}\,\,||\,\vec{u}] \) for any particular nonnegative value of \(\alpha\). However, the reverse statement is not true. It is important to note that a single order-\(\alpha\) R\'{e}nyi divergence does not establish the majorization pre-order. For \(\vec{q} \prec \vec{p}\) to hold, it is necessary that \(S_\alpha[\vec{q}\,\,||\,\vec{u}] \leq S_\alpha[\vec{p}\,\,||\,\vec{u}] \) for all nonnegative values of \(\alpha\). On the other hand, when \(\alpha\) approaches to 1 the limit gives the Kullback–Leibler divergence, which is sufficient to imply the transformation law in conventional thermodynamics. \(S_1[\vec{x}\,||\,\vec{y}]\), also known as relative entropy is defined as \(\sum_i p_i \log p_i/q_i\).

In this review, our focus has been solely on deterministic state transformations of single-copy systems. For asymptotic, probabilistic, catalytic state transformations, and more, we refer the reader to Ref.~\cite{2015_PhysReport_Nicole_Review}. Additionally, we have not addressed all the nonuniformity monotones here. A comprehensive summary of these monotones can also be found in Tables 1, 2, and 3 in Ref.~\cite{2015_PhysReport_Nicole_Review}.


\subsection{Resource Theory of Athermality}\label{Sec:Athermality}

Now we can move on from isolated systems to closed systems that can exchange both information and energy with their surroundings. According to conventional thermodynamics, the transition from one equilibrium state to another in these systems is possible when there is a decrease in free energy. The free energy, which can be expressed in the form of \(F = U - T \, S\), is dependent on both the entropy of the system and its average energy \(U \equiv \text{tr}[\rho \, \hat{H}]\). Here, \(\hat{H}\) represents the Hamiltonian of the system, and \(T\) is the temperature of the thermal environment with which it is in equilibrium. When \(\hat{H}\) vanishes or \(T\) goes to infinity, the contribution from the average energy can be neglected, and as discussed in the previous section, entropy takes the place of free energy.

Based on our extensive examination of various resources illustrated in Fig.~\ref{FIG:Sum}, it has become evident that the utilization of majorization-based criteria is imperative, rather than relying on entropy, to establish a hierarchical order among states based on their resource contents. The fundamental motivation behind this choice stems from the necessity to transcend the thermodynamic limit, commonly known as identical and independent distributions (iid), when investigating quantum systems. Unlike classical systems, quantum systems cannot be replicated when their state is unknown. However, in the derivation of thermodynamic laws, it is conventional to assume the availability of numerous identical copies of a system. Therefore, when we surpass the iid limit, the following question arises: What can we substitute for free energy? In what follows, we will explore how majorization can also provide assistance in this context. However, prior to delving into that, let us first clarify what resources are considered relevant for agents who have access to a thermal bath at a constant temperature, as well as those that are not.

The only state that can be obtained freely through thermal operations (TOs) is the Gibbs state. It is represented by the equation:
\begin{equation}
    \varrho_{S|\beta} = \frac{1}{{\mathcal{Z}_{S|\beta}}} e^{-\beta \hat{H}_S}.
\end{equation}
Here, \(\hat{H}_S\) denotes the Hamiltonian of the system, \(\mathcal{Z}_{S|\beta}\) is the partition function, which is equal to the trace of \(e^{-\beta \hat{H}_S}\), and \(\beta = 1/{k_B T}\) represents the inverse temperature of the environment, with \(k_B\) being the Boltzmann constant. \(\varrho_{S|\beta}\) is represented by a diagonal matrix that obeys the Boltzmann distribution at inverse temperature \(\beta\), i.e., \(\varrho_{S|\beta} = diag(\vec{\gamma}_{S|\beta})\) where \(\vec{\gamma}_{S|\beta} = \{e^{-\beta E_j^S}/\mathcal{Z}_{S|\beta}\}\) such that \(\hat{H}_S = \sum_j E_j^S \ketbra{j}{j}\). Any states other than it, even Gibbs states at different temperatures, are considered out-of-equilibrium. These nonequilibrium states are valuable as resource states in the context of thermal operations (TOs). The degree to which these states differ from thermal equilibrium at inverse temperature \(\beta\) is referred to as their athermality, which is the primary focus and resource of interest in this context.

Considering a quantum system with a state \(\rho\) and a Hamiltonian \(\hat{H}\), TOs are CPTP maps \(\mathcal{T}: \mathcal{L}(\mathcal{H}) \rightarrow \mathcal{L}(\mathcal{H}^\prime)\) that can be expressed as follows~\cite{2000_IntJTheorPhys_TOs, Brandao2013RTofThermo}:
\begin{equation} \label{Eq::TOs}
\bigg\{\rho \mapsto \rho^\prime= \mathrm{tr}_A \left[U (\rho \otimes \varrho_{B|\beta}) U^\dagger\right], \quad \hat{H} \mapsto \hat{H}^\prime = \mathrm{tr}_A \left[\hat{H}_{tot} - \hat{H}_A\right] \bigg\}.
\end{equation}
In this equation, \(\varrho_{B|\beta}\) represents the thermal equilibrium state of an ancillary system \(B\), given by \(e^{-\beta \hat{H}_B}/\mathcal{Z}_{B|\beta}\), where \(\hat{H}_B\) is the Hamiltonian of system \(B\) and \(\mathcal{Z}_{B|\beta}\) is the corresponding partition function. The unitary operation \(U\) acts on the joint system and preserves the total energy, i.e., \([\hat{H}_{tot}, U] = 0\) where \(\hat{H}_{tot} = \hat{H} \otimes \mathbb{1}_B + \mathbb{1} \otimes \hat{H}_B\). The process explained in Eq.~\eqref{Eq::TOs} can be broken down into three steps. Initially, an agent connects the main system with an ancillary system \(B\) that is in thermal equilibrium at temperature \(T\). This enables the exchange of heat between the two systems while keeping the total entropy and total energy unchanged. Subsequently, the agent traces outs an arbitrary subsystem \(A\) from the entire setup.

There are alternative approaches to RT of athermality, such as thermal processes~\cite{2015_PRL_CohinRTD, 2015_NatCommun_qCohinTD, 2015_PRX_KK_ThermalCoherence, 2018_NatureComm_qMajor} that can be defined by two key properties. Firstly, the Gibbs state at temperature \(T\) should remain unchanged under operations that do not require any work, i.e., \(\mathcal{T}(\varrho_{S|\beta}) = \varrho_{S|\beta}\). Secondly, the creation of any quantum coherence in the energy basis cannot occur without cost, which is analogous to having symmetry under time translations, i.e., \(\mathcal{T}(e^{- i \hat{H}_S t} \rho_S \, e^{i \hat{H}_S t}) = e^{- i \hat{H}_S t} \, \mathcal{T}(\rho_S) \, e^{i \hat{H}_S t}\). The question of whether the set of states achievable through thermal processes aligns with the set of states achievable by TOs remains an intriguing open question~\cite{2015_NJP_GibbsPrervingMapsVsTOs, 2018_Quantum_eTOs}. A recent study~\cite{2023_arXiv_KK_TOandTP} has shed light on this matter, revealing that energy-incoherent states achievable by TOs can be closely approximated through memory-assisted Markovian thermal processes, provided a sufficiently large memory is available. This framework presents a promising avenue for advancing our understanding of the role played by memory in diverse thermodynamic scenarios.

If both the initial and target states lack energetic coherence, we can determine the feasibility of transforming the state \(\rho\) into \(\sigma\) through the utilization of a TO. Let us assume that \(\rho = \sum_j p_j \ket{j}\bra{j}\) and \(\sigma = \sum_j q_j \ket{j}\bra{j}\), where \(\{E^S_j, \ket{j}\}\) forms the eigenspectrum of \(\hat{H}_S\). In the presence of a TO at inverse temperature \(\beta\), which converts \(\rho\) into \(\sigma\), there should be a relationship between the probability distributions \(\vec{p}\) and \(\vec{q}\) given by \(G^{\mathcal{T}} \vec{p} = \vec{q}\), where \(G^{\mathcal{T}}\) represents a Gibbs-stochastic matrix, a stochastic map that preserves the Boltzmann distribution \(\vec{\gamma}_{S|\beta}\).

Now, we can extend the majorization criterion to include finite temperatures and non-trivial Hamiltonians. This extension enables us to introduce a nonequilibrium generalization of free energy, much like how the previous section enhanced our comprehension of entropy beyond iid limit. Let us assume the existence of an embedding map, denoted as \(\Gamma_d\), which is known as the Gibbs-rescaling map. This map transforms the Boltzmann distribution \(\vec{\gamma}_{S|\beta}\) with \(n\) rational elements into a uniform distribution \(D\) where \((\gamma_{S|\beta})_j \approx d_j/D\). Then, there exists a Gibbs-stochastic matrix \(G^{\mathcal{T}}\) that converts \(\vec{p}\) into \(\vec{q}\) iff \(\Gamma_d(\vec{p})\) majorizes \(\Gamma_d(\vec{q})\). We can express this condition as the so-called \textit{thermomajorization criterion}~\cite{2013_NatureComm_HorodeckiOppenheim} that exclusively involves \(\vec{p}\) into \(\vec{q}\) as follows:
\begin{eqnarray}\label{Def:ThermoMajorization}
\sum_{i=1}^k p_i^{\downarrow \beta} e^{\beta E^S_{i}} \geq \sum_{i=1}^k q_i^{\downarrow \beta} e^{\beta E^S_{i}} \quad \text{for all} \;  k=1, 2, \dots, n,
\end{eqnarray}
where \(x_i^{\downarrow \beta}\) is called \(\beta\)-ordering of \(x_i\) defined by \(x_{\pi(i)}\) with \(\pi\) is the permutation ensuring \(x_{i+1} \, e^{\beta E^S_{i+1}} \geq x_{i} \, e^{\beta E^S_{i}}\).

The concept of thermomajorization introduces a set of inequalities that is the necessary and sufficient condition for thermodynamic state transformations in various systems, including microscopic, out-of-equilibrium, and highly correlated ones. In the presence of an ancillary system that returns back into its original state with a probability close to unity, the application of the thermomajorization criterion to the entire system establishes a family of second laws for the system of interest~\cite{Brando2015TheSecondLaw}. By changing the energy levels of a system and facilitating thermalization between any two energy levels within it, while utilizing an ancilla in a thermal state, it becomes possible to transform any state into another state that adheres to these second laws~\cite{2018_PRX_SufficientTOs}. Importantly, these second laws converge to the conventional second law of thermodynamics in the iid limit, as well as in scenarios where the ancillary system becomes correlated with the system of interest while still maintaining its own state intact \cite{2018_PRX_Muller_CorrelatedCatalyzerAndUniqueF}.

In the aforementioned family of second laws, free energy is replaced by athermality monotones that are called \(\alpha\)-free energies~\cite{2013_NatureComm_HorodeckiOppenheim, 2013_NatureComm_Aberg, 2015_NJOfPhys_Oscar1, 2015_NJOfPhys_Oscar2} and are defined by
\begin{eqnarray}
    F_\alpha(\vec{x}) = - k_B T \log \mathcal{Z}_{S|\beta} + k_B T S_\alpha[\vec{x}\,\,||\,\vec{\gamma}_{S|\beta}] ,
\end{eqnarray}
where \(S_\alpha\) are \(\alpha\)-R\'{e}nyi divergences defined in Eq.~\eqref{Eq::RenyiDiv}. As \(S_1[\vec{x}\,\,||\,\vec{\gamma}_{S|\beta}]\) equals to \(\beta \, U - S(\vec{x}) + \log \mathcal{Z}_{S|\beta}\), we will end up with the conventional free energy, \(F = U - T \, S\) in the limit \(\alpha\) goes to 1.

RT of thermality offers us a framework based on stochastic maps and majorization-based partial orders, as summarized. Within this framework, there are many other concepts that we have not addressed in this review but are useful in various different areas, such as relative thermalization~\cite{2016_PRE_RennerEtAll_RelThermalization}, conditioned thermal operations~\cite{2017_PRA_Gour_ConditionedTOs}, relative submajorization~\cite{2016_Renes_RelSubMajorization}, continuous thermomajorization~\cite{2022_PRA_KK_ContThermoMaj, 2022_PRL_KK_OptimizeMaj, Plenio_PRA_2022}, quantum majorization~\cite{2018_NatureComm_qMajor} and thermal cones~\cite{2017_PRA_KK_TDArrowofTime, 2022_PRE_KK_ThermalCone}.

Thus far, our focus has been solely on the realm of RT of athermality in the single-shot limit. For situations that involve asymptotic, probabilistic, catalytic state transformations, and other related scenarios,  we recommend referring to the pedagogical review written by Lostaglio~\cite{2019_RepProgPhys_Lostaglio}. Additionally, we have assumed in this review that the heat baths accessible to agents have infinitely many degrees of freedom. The alterations in state transformations achievable through NOs and TOs in the presence of finite heat baths can be found in Ref.~\cite{Scharlau2018quantumhornslemma}


\subsection{Resource Theory of Nonequilibrium}\label{Sec:Nonequilibrium}

Finally, the range of quantum operations that agents can perform on a system can be expanded by considering the exchange of various quantities, such as information, heat, particles, and more. Within the framework of equilibrium thermodynamics, the state transformations of these open systems are characterized by thermodynamic potentials, including the grand potential \(\Phi\), which can be expressed as \(U - T \, S - \sum_j \mu_j Q^{(j)}\) where \(\mu_j\) corresponds to the chemical potential associated with the extensive quantity \(Q^{(j)}\).

In a similar manner to our transition from RT of nonuniformity to RT of athermality in the preceding section, we can further expand the scope of thermodynamic resource theories to encompass domains beyond heat baths~\cite{Nicole_2016_PRE_equiMaj, 2018_Nicole_JPA_BeyondHeatBaths}. In the case of an agent having access to generalized thermodynamic baths, systems lacking resource value will be in the equilibrium state given by the density operator
\begin{equation}
    \varrho = e^{-\beta \hat{H} + \sum_j \mu_j \hat{Q}^{(j)}}/\mathcal{Z} ,
\end{equation}
where \(\mathcal{Z}\) is the partition function. Free operations in this context will take the form
\begin{equation}
    \bigg\{\rho \mapsto \rho^\prime = \mathrm{tr}_A \left[U (\rho \otimes \varrho_B) U^\dagger\right], \;  \hat{H} \mapsto \hat{H}^\prime = \mathrm{tr}_A \left[\hat{H}_{tot} - \hat{H}_A\right], \;  \hat{Q}^{(j)} \mapsto \hat{Q^\prime}^{(j)} = \mathrm{tr}_A \left[\hat{Q}^{(j)}_{tot} - \hat{Q}^{(j)}_A\right] \bigg\} ,
\end{equation}
with the conditions \(\left[\hat{H}_{tot}, U\right] = \left[\hat{Q}^{(j)}_{tot}, U\right] = 0\). We can consider these free operations, called equilibrating operations, in a similar manner to noisy and thermal operations, through three consecutive steps. The crucial aspect here is the preservation of the total amount of thermodynamic charges during the interaction of the system with the ancillary system \(B\).

To analyze general non-equilibrium states based on their resource contents, we can adopt the same method that extends majorization to thermomajorization. When all thermodynamic charges are commutative, the free states can be represented by diagonal matrices, simplifying the problem to probability distributions. By identifying the stochastic matrix that preserves the equilibrium probability distribution, we can then apply an embedding map to transform this stochastic matrix into doubly stochastic matrices. This map allows us to generalize the concept of majorization to \textit{equimajorization}~\cite{Nicole_2016_PRE_equiMaj}, which has previously been referred to as \textit{d-majorization}~\cite{dMaj} or mixing distance~\cite{MixingDistance1, MixingDistance2} in the literature.

The concept of equimajorization allows us to generalize nonequilibrium monotones. However, the assumption of commutativity of thermodynamic charges, which is necessary for this generalization, is challenged by quantum phenomena like uncertainty relations that demonstrate the non-commutativity of observables. The impact of noncommuting charges on thermodynamic phenomena has become a topic of interest at the intersection of quantum information theory and thermodynamics~\cite{2016_Guryanova_NatCommun, 2016_Nicole_NatCommun,2017_NJP_Lostaglio}, particularly in the field of many-body physics~\cite{2020_Nicole_PRE_NoncommutingCharges, 2023_Nicole_PRB_NoncommutingCharges, 2023_Nicole_PRXQ_NoncommutingCharges}. Recent advancements in this area have unveiled significant consequences arising from the noncommutation of charges~\cite{2018_Nicole_JPA_BeyondHeatBaths, PRXQuantum.3.010304, PhysRevLett.130.140402}. For further exploration of this captivating subject, we recommend referring to the perspective provided in Ref.~\cite{2023_Nicole_arXiv_PerspectiveRTNN}.


\section{Outlook}\label{Sec:Conclusion}

The framework of QRTs \cite{Chitambar2019-QRTs}  allows for a systematic approach to investigate, quantify, and manipulate distinct quantum resources. QRT can be used not only to analyze the limits and potential of existing quantum devices but also to design quantum-enhanced devices and protocols. By understanding the properties and transformations of quantum resources, researchers can harness these resources to develop novel strategies that outperform classical counterparts or improve the performance of existing quantum technologies. In this concise yet comprehensive review, we have initiated our discourse by presenting the fundamental constituents that underpin the construction of any QRT. Subsequently, a succinct introduction to the concept of majorization has been presented, highlighting its fundamental significance in the QRTs expounded upon.

Then, within the second part of this review, we have presented a concise assessment of the RTs of bipartite quantum entanglement \cite{Horodecki2009-QEnt}, quantum coherence \cite{Baumgratz2014QCoherence}, and superposition \cite{Theurer2017RTS}. The majorization-based transformation procedures for all three RTs are well-established in the context of pure initial and target resource states. However, extending these transformation methods to incorporate mixed states poses a significant and unsolved challenge. This stems from the inherent difficulty in fully harnessing the mentioned resources within existing QTs without complete isolation from their surrounding environment.

An ongoing challenge in the field of entanglement theory is finding a solution for handling multipartite entanglement within the RT framework. A comprehensive RT of multipartite entanglement is crucial for scaling up quantum devices and enabling the utilization of entanglement as a resource from micro and nano scales to meso and macro scales. One of the primary complexities involved in tackling this issue is the existence of various classes of genuine entanglement in the multipartite regime. To illustrate, in tripartite entanglement, GHZ-type and W-type entanglements should be regarded as separate resources (i.e., contain different types of entanglement), each governed by distinct constraints imposed on LOCCs, resulting in different sets of free operations.

Although quantum coherence is a special case of quantum superposition as shown in Fig.~\ref{FIG:Sum}, their unification within the framework of RT remains an active area of research. While coherence has been extensively studied and reviewed \cite{Streltsov2017-RTofCoh}, the study of superposition \cite{Theurer2017RTS} is relatively new and offers numerous opportunities for further investigation. One significant challenge in quantifying and manipulating superposition as a resource arises from the complexity introduced by nonorthogonality. By adopting the biorthogonality framework introduced in Ref.~\cite{pusuluk2022unified} and exploring the relationship between coherence and superposition as in Ref.~\cite{Torun2023Lowdin}, we can strive towards a complete RT of nonclassicality that treats superposition, coherence, and correlations on equal footing. Within this context, investigating superposition within the resource-theoretic framework represents a promising frontier, with the potential to reveal unique characteristics and practical implications in the field of QTs. Sec.~\ref{Sec:RTSuperposition} aims to provide a rationale for discussing the previously introduced RTs and outlines potential avenues for future research on quantum superposition.

In the final part of this review, we have delved into the RTs of nonequilibrium thermodynamics, drawing structural parallelism with the RTs of quantum information discussed earlier. For instance, just as the nonorthogonality of states leads to a complexity in the RTs of quantum information, the non-commutativity of observables similarly gives rise to a comparable complexity. Additionally, the majorization-based transformation procedures we have summarized were limited to pure states in Sec.~\ref{Sec:Information-theoretic}, while they were restricted to mixed states in Sec.~\ref{Sec:ThermodynamicsApproach}. However, significant efforts have been dedicated to addressing these limitations in the RTs of nonequilibrium thermodynamics. Particularly, extensive research has been conducted in the literature on the changes in energetic coherence under thermal operations, exploring the realm of the RT of athermality \cite{2015_PRL_CohinRTD, 2015_NatCommun_qCohinTD, 2015_PRX_KK_ThermalCoherence, 2018_NatureComm_qMajor}.

In the case of nondegenerate energy levels, the energetic populations and coherences vary independently under thermal operations. However, in degenerate systems, interconversions between populations and coherence can take place. Coherence types that enable such interconversions are classified as ``heat-exchange'' \cite{2016_Entropy_Ozgur, 2019_PRE_Ozgur_Multiatom, pusuluk2021prr}, ``internal'' \cite{AHF_2019_PhysRevResearch_Petruccione} or ``horizontal'' \cite{AutoThermalMach_2019, 2020_PRA_HorizantolCohAndPops, 2020_arXiv_2006_01166} in the field of open quantum system dynamics theory. Extensive research has been conducted to investigate the role of this specific coherence type in thermal processes using the master equation. In the framework of RT, the same coherence type is referred to as mode-zero coherence \cite{2019_RepProgPhys_Lostaglio},  but its potential has not been adequately examined through majorization-based investigations in the existing literature.

To conclude, quantum protocols that utilize multiple types of resources simultaneously have the potential to demonstrate superior performance compared to those relying on a uniform quantum fuel \cite{2023_arXiv_KK_ResourceEngines}. With this motivation in mind, the objective of this review is to highlight the structural similarities among various quantum resources in RTs. As a result, we were unable to provide an exhaustive literature summary for any specific QRT. Instead, our aim has been to direct the reader toward other reviews in the field to the best of our ability. We expect that this compendious review will contribute to unveiling the commonalities among various RTs, thereby enhancing our comprehension of genuine quantum phenomena in technology.


\section*{Acknowledgments}\label{Sec:Acknowledgment}
\addcontentsline{toc}{section}{Acknowledgment} 

G. Torun acknowledges financial support from TÜBİTAK Research Institute for Fundamental Sciences. O. Pusuluk acknowledges financial support from the Scientific and Technological
Research Council of Türkiye (TÜBİTAK) under Grant No.~120F089.



%

\addcontentsline{toc}{section}{References} 


\end{document}